\documentclass[amssymb,prb,twocolumn,superscriptaddress,floats,showkeys,amsmath]{revtex4-1}

\usepackage{graphicx}
\usepackage{epstopdf}
\usepackage{hyperref} 
\hypersetup{colorlinks,citecolor=blue, filecolor=blue ,linkcolor=blue , urlcolor=blue, pdftex}

\usepackage[labeled,resetlabels]{multibib}

\newcites{SI}{SupI}

\begin{document}

\title{Energy spectrum of two-dimensional excitons in a non-uniform dielectric medium}

\author{M. R. Molas}
\email{maciej.molas@fuw.edu.pl}
\affiliation{Laboratoire National des Champs Magn\'etiques Intenses, CNRS-UGA-UPS-INSA-EMFL, 25, avenue des Martyrs, 38042 Grenoble, France}
\affiliation{Institute of Experimental Physics, Faculty of Physics, University of Warsaw, ul. Pasteura 5, 02-093 Warszawa, Poland}
\author{A. O. Slobodeniuk}
\affiliation{Laboratoire National des Champs Magn\'etiques Intenses, CNRS-UGA-UPS-INSA-EMFL, 25, avenue des Martyrs, 38042 Grenoble, France}
\author{K. Nogajewski}
\affiliation{Laboratoire National des Champs Magn\'etiques Intenses, CNRS-UGA-UPS-INSA-EMFL, 25, avenue des Martyrs, 38042 Grenoble, France}
\affiliation{Institute of Experimental Physics, Faculty of Physics, University of Warsaw, ul. Pasteura 5, 02-093 Warszawa, Poland}
\author{M.~Bartos}
\affiliation{Laboratoire National des Champs Magn\'etiques Intenses, CNRS-UGA-UPS-INSA-EMFL, 25, avenue des Martyrs, 38042 Grenoble, France}
\affiliation{Central European Institute of Technology, Brno University of Technology,  Purky\v{n}ova 656/123, 612 00 BRNO, Czech Republic}
\author{\L{}. Bala}
\affiliation{Laboratoire National des Champs Magn\'etiques Intenses, CNRS-UGA-UPS-INSA-EMFL, 25, avenue des Martyrs, 38042 Grenoble, France}
\affiliation{Institute of Experimental Physics, Faculty of Physics, University of Warsaw, ul. Pasteura 5, 02-093 Warszawa, Poland}
\author{A. Babi\'nski}
\affiliation{Institute of Experimental Physics, Faculty of Physics, University of Warsaw, ul. Pasteura 5, 02-093 Warszawa, Poland}
\author{K. Watanabe}
\affiliation{National Institute for Materials Science, 1-1 Namiki, Tsukuba 305-0044, Japan}
\author{T. Taniguchi}
\affiliation{National Institute for Materials Science, 1-1 Namiki, Tsukuba 305-0044, Japan}
\author{C. Faugeras}
\affiliation{Laboratoire National des Champs Magn\'etiques Intenses, CNRS-UGA-UPS-INSA-EMFL, 25, avenue des Martyrs, 38042 Grenoble, France}
\author{M. Potemski}
\email{marek.potemski@lncmi.cnrs.fr}
\affiliation{Laboratoire National des Champs Magn\'etiques Intenses, CNRS-UGA-UPS-INSA-EMFL, 25, avenue des Martyrs, 38042 Grenoble, France}
\affiliation{Institute of Experimental Physics, Faculty of Physics, University of Warsaw, ul. Pasteura 5, 02-093 Warszawa, Poland}

\begin{abstract}
We demonstrate that, in monolayers (MLs) of semiconducting transition metal dichalcogenides, the $s$-type Rydberg series of excitonic states
 follows a simple energy ladder: $\epsilon_n=-Ry^*/(n+\delta)^2$, $n$=1,2,\ldots, in which $Ry^*$ is very close to the Rydberg energy scaled by the dielectric
 constant of the medium surrounding the ML and by the reduced effective electron-hole mass, whereas the ML polarizability is only accounted
 for by $\delta$. This is justified by the analysis of experimental data on excitonic resonances, as extracted from magneto-optical measurements
 of a high-quality WSe$_2$ ML encapsulated in hexagonal boron nitride (hBN), and well reproduced with an analytically solvable Schr\"odinger
 equation when approximating the electron-hole potential in the form of a modified Kratzer potential. Applying our convention to other,
 MoSe$_2$, WS$_2$, MoS$_2$ MLs encapsulated in hBN,  we estimate an apparent magnitude of $\delta$ for each of the studied structures.
 Intriguingly, $\delta$ is found to be close to zero for WSe$_2$ as well as for MoS$_2$ monolayers, what implies that the energy ladder of
 excitonic states in these two-dimensional structures resembles that of Rydberg states of a three-dimensional hydrogen atom.
\end{abstract}

\maketitle

Coulomb interaction in a non-uniform dielectric medium~~\cite{rytova,keldysh}, is one of the central points in
investigations of large classes of nanoscale materials, such as, for example, graphene~\cite{kotov2012,faugeras2015}
and other atomically thin crystals including their heterostructures~\cite{Geim2013}, as well as colloidal
nanoplatelets~\cite{Ithurria2011}, and two-dimensional perovskites~\cite{Stoumpos2016,Blancon2018}. This problem has been, in
recent years, particularly largely discussed in reference to a vast amount of investigations of excitons in monolayers (MLs)
of semiconducting transition metal dichalcogenides \mbox{(S-TMDs)}~\cite{chernikov,Raja2017,koperski,Wang2018,goerbig2018}. Surprisingly,
at first sight, the Rydberg series of $s$-type excitonic states in these archetypes of two-dimensional (2D) semiconductors,
does not follow the model system of a 2D hydrogen atom~\cite{macdonald,koteles,christol}, with its characteristic energy
sequence, \mbox{$\sim1/(n-1/2)^2$}, of states with a principal quantum number $n$. The main reason for that is a dielectric inhomogeneity
of the 2D S-TMD structures, $i.e.$, MLs surrounded by (deposited/encapsulated on/in) alien dielectrics. At large
electron-hole ($e$-$h$) distances, the Coulomb interaction scales with the dielectric response of the surrounding medium whereas
it appears to be significantly weakened at short $e$-$h$ distances by the usually stronger dielectric screening in the 2D plane.
A common approach to account for the excitonic spectra of S-TMD MLs refers to the numerical solutions of the
Schr\"odinger equation, in which the $e$-$h$ attraction is approximated by the Rytova-Keldysh (RK) potential~\cite{rytova,keldysh}.
The RK approach has been used to explain a number of excitonic features in S-TMD MLs~\cite{stierhBN}. However, it is only solvable
numerically. A more phenomenological and intuitive approach, presented below, might be an optional solution
to this problem.

In this Letter, we demonstrate that the energy spectrum, $\epsilon_n$ ($n$=1, 2,\dots), of Rydberg series of $s$-type excitonic
states in S-TMD MLs may follow a simple energy ladder: $\epsilon_n=-Ry^*/(n+\delta)^2$. From magneto-optical investigations
of a WSe$_2$ ML encapsulated in hexagonal boron nitride (hBN), we accurately establish that \mbox{$Ry^*$=140.5~meV} and
$\delta$=-0.083 in this particular S-TMD system. The $\epsilon_n$ spectrum, with $\delta\sim0$, turns out to closely
reflect the characteristic spectrum of a three-dimensional (3D) hydrogen atom. The $\epsilon_n=-Ry^*/(n+\delta)^2$ ansatz is well reproduced
with an analytical theoretical approach in which the $e$-$h$ potential is assumed to have the form of a modified Kratzer
potential~\cite{kratzer}. $Ry^*$ is identified with the effective (3D) Rydberg energy, 
$Ry\times\mu/(\varepsilon^2m_0)$, scaled
by the dielectric constant $\varepsilon$ of the surrounding hBN medium and the reduced $e$-$h$ mass $\mu=(m_em_h)/(m_e+m_h)$,
with $Ry$=13.6~eV; $m_e$ and $m_h$ are, correspondingly, the electron and hole effective masses, and $m_0$ is the free
electron mass. Dispersion of $Ry^*$ and $\delta$ parameters in different studied samples, WSe$_2$, MoSe$_2$, MoS$_2$,
and WS$_2$ MLs encapsulated in hBN, is discussed and the reduced $e$-$h$ masses in these ML structures are
estimated.

\begin{figure*}[t]
    \centering
    \includegraphics[width=1\linewidth]{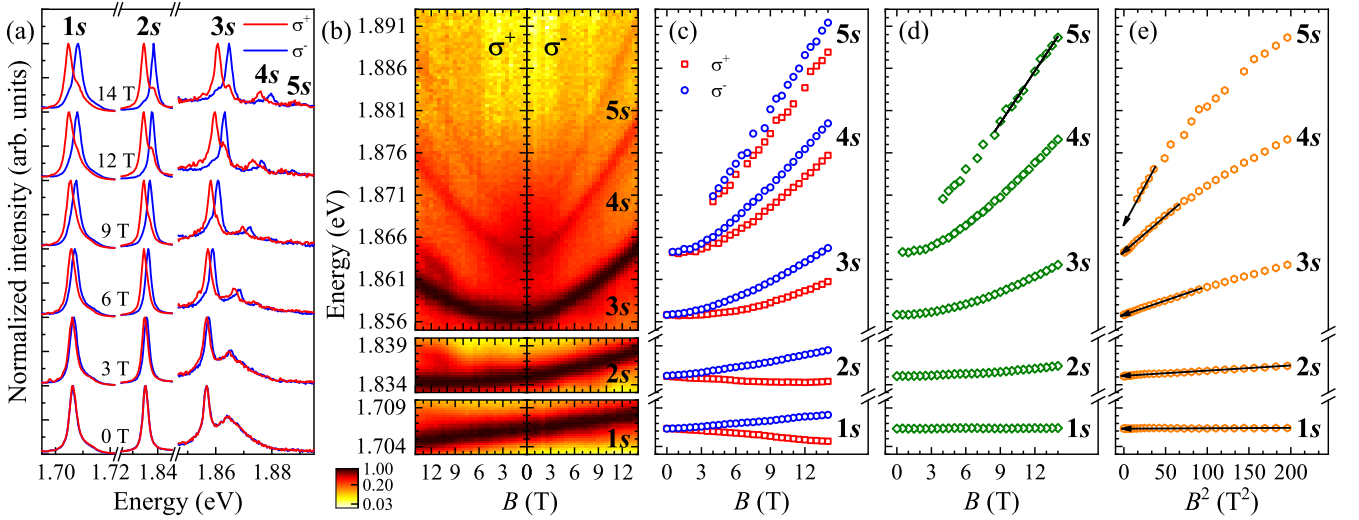}%
    \caption{(a) Helicity-resolved ($\sigma^{\pm}$) PL spectra of WSe$_2$ ML at selected magnetic fields. The separate parts of the
    spectra are normalized to the intensity of the 1$s$, 2$s$, and 3$s$ lines. (b) False-colour map of the corresponding PL spectra
    from 0 to 14~T. (c) Obtained excitonic energies for $\sigma^{\pm}$ components as a function of magnetic fields. Mean energies of
    the $\sigma^{\pm}$ components of excitonic resonances measured on WSe$_2$ ML as a function of (d) $B$ and (e) $B^2$. The black
    lines are obtained by fitting the presented data with (d) $E_{5s}(B)=A+9/2\hbar \omega^*_c$ ($A$ is a fitting parameter) and (e)
    $E_{ns}(B)=E_{ns}(B=0)+\sigma B^2$.}
    \label{fig:fig_1}
\end{figure*}

To accurately determine the characteristic ladder of
\mbox{$s$-type} excitonic resonances in the experiment, we
profited of a particularly suitable for this purpose method of
magneto-optical spectroscopy~\cite{akimoto,potemski1991}. The
active part of the structure used for these experiments was a
WSe$_2$ ML embedded in between hBN layers. More details on
samples' preparation and on the experimental techniques can be
found in the Supplemental Materials (SM). We measured the
(circular) polarization resolved magneto-photoluminescence (PL) at
low temperatures (4.2~K) and in magnetic fields up to 14~T,
applied in the direction perpendicular to the monolayer plane.
Here we focus on magneto-PL spectra of our WSe$_2$ ML, observed in
the spectral range from $\sim$1.7 to $\sim$1.9~eV. As shown in
Fig.~\ref{fig:fig_1}(a) and (b), these spectra are composed of up
to five PL peaks, which are clearly resolved in the range of high
magnetic fields. Following a number of previous
investigations~\cite{manca2017,Chow2017,chen2017,stierhBN,Liu2019}
on similar structures, the observed PL peaks are identified with a
series of excitonic resonances forming the 1$s$, 2$s$, \dots, 5$s$
Rydberg series of the so-called A
exciton~\cite{koperski,Wang2018}. Each $ns$ PL peak, $n$=1, 2,
\dots, 5, demonstrates the valley Zeeman effect. This is
illustrated in Fig.~\ref{fig:fig_1}(c) in which the energies of
the $\sigma^+$ and $\sigma^-$—polarized PL peaks are plotted as
a function of the magnetic field. In accordance with previous
reports we extract \mbox{$g$=-4.1} for valley $g$-factor of the
1$s$ resonance, but read a significantly stronger valley Zeeman
effect for all excited states ($g\sim$-4.8). The later observation
is intriguing and should be investigated in more details, which
is, however, beyond the scope of the present paper.
We have also carried out the magneto-PL
experiments on MoS$_2$ and WS$_2$ monolayers encapsulated in hBN,
but only the 1$s$ and 2$s$ resonances could be observed in these
structures in the range of magnetic fields applied (see SM for
details).

The magnetic field evolution of the mean energies of $\sigma^+$
and $\sigma^-$ PL peaks is illustrated in Figs~\ref{fig:fig_1}(d)
and (e). These energies, $E_{ns}$, are plotted as a function of
the magnetic field $B$ in Fig.~\ref{fig:fig_1}(d), and as a
function of $B^2$ in Fig.~\ref{fig:fig_1}(e), which illustrates
the characteristic but distinct behavior of $ns$ states in the
so-called low- and high-field
regime~\cite{macdonald,potemski1991}. The high field limit, for a
given $ns$ resonance, appears when $l_B\ll r_{ns}$, or conversely
when binding energy of the $ns$ state $E^{ns}_b\ll \hbar
\omega^*_c/2$. Here $r_{ns}$ and $E^{ns}_b$ denote,
correspondingly, the mean lateral extension and the binding energy
$E^{ns}_b$=$E_g$-$E_{ns}$ of a given $ns$ state at $B$=0, $\hbar
\omega^*_c=\hbar e B/\mu$, $l_B=\sqrt{\hbar/e B}$
is the magnetic length and other symbols have their conventional meaning.
In the high-field limit, the energies of $E_{ns}$
resonances approach a linear dependence upon $B$, with a slope
given by $(n-1/2)\hbar \omega^*_c$. In the low field limit
($l_B\gg r_{ns}$, $E^{ns}_b\gg\hbar\omega^*_c/2$), the $ns$
resonances display the diamagnetic shifts:
$E_{ns}(B)=E_{ns}(B=0)+\sigma B^2$, where $\sigma=(e r_{ns})^2/8
\mu$ is the diamagnetic coefficient. The 1$s$ and 2$s$ resonances
follow the low-field regime in the entire range of the magnetic
fields investigated due to their small exciton's
radii and/or large binding energies, see Fig.~\ref{fig:fig_1}(e).
The high field regime is approached for the 5$s$ resonances with
an approximate linear dependence of $E_{5s}$ with $B$, in the
range above $\sim$8~T. This linear dependence, marked with a solid
line in Fig.~\ref{fig:fig_1}(d), displays a slope of 2.1~meV/T,
which if compared to $(9/2)\hbar \omega^*_c$ dependence, provides an
estimate of 0.25~$m_0$ for the reduced mass in the WSe$_2$ ML.
However, one may also argue that working with magnetic fields up
to 14~T only, the high field limit is still barely developed even
for the 5$s$ state. In this context, our estimation of the reduced
effective mass should be seen as its upper bound and, in the
following we assume $\mu$=0.2~$m_0$ for the WSe$_2$ ML, following
the results of experiments performed in fields up to
60~T~\cite{stierhBN}.

\begin{figure}[t]
    \centering
    \includegraphics[width=0.7\linewidth]{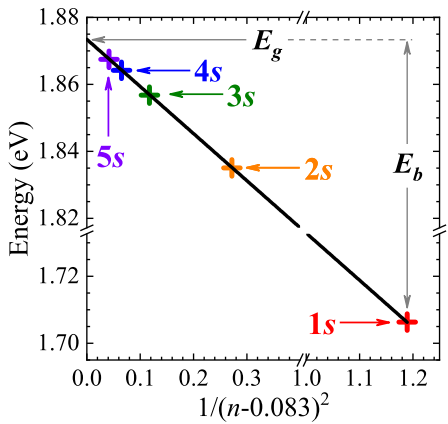}%
    \caption{Experimentally obtained transition energies for the exciton states as a function $1/(n+\delta)^2$ for $\delta$=-0.083.
    The black line shows a fit of the data to the model described by Eq.~\ref{eq:eq_1}. The grey lines denote the band-gap ($E_g$)
    and excitonic binding ($E_b$) energies.}
    \label{fig:fig_3}
\end{figure}

In the following we focus on the energy sequence $E_{ns}$ of 1$s$, 2$s$, \dots, 5$s$ excitonic resonances as they appear in the
absence of magnetic field. As shown in Fig.~\ref{fig:fig_1}(e), the apparent $E_{ns}$ values are accurately determined with linear
extrapolations of $E_{ns}$ versus $B^2$ dependences to B=0. Next, we put forward a hypothesis that the energy sequence $E_{ns}$ obeys
the following rule:
\begin{equation}
E_{ns}=E_g-\frac{Ry^*}{(n+\delta)^2},
\label{eq:eq_1}
\end{equation}
where, at this point, $E_g$, $Ry^*$, and $\delta$ should be regarded as unknown adjustable parameters. To test the above formula
against experimental data, we note that Eq.~\ref{eq:eq_1} implies that (for example) the ratio ($E_{3s}-E_{1s})/(E_{2s}-E_{1s})$ only depends
on $\delta$, and, reading this ratio from the experiment, we extract $\delta$=-0.083. With this value we find (see Fig.~\ref{fig:fig_3})
that our experimental $E_{ns}$ series perfectly matches Eq.~\ref{eq:eq_1}, and, at the same time, we determine two other parameters,
$E_g$=1.873~eV and $Ry^*$=140.5~meV (or conversely, exciton binding energy $E_b=E_g-E_{1s}=Ry^*/(1-0.083)^2$=167~meV). The above
$E_g$ and $E_b$ values are in very good agreement with those already reported in the literature~\cite{stierhBN}. Relevant for
our further analysis, is the observation that the derived value for $Ry^*$ coincides well with the effective (3D) Rydberg energy
$Ry^*$=13.6~eV$\cdot\mu/(\varepsilon^2m_0$)=134.3~meV, scaled by the dielectric constant of the surrounding hBN material
$\varepsilon=\varepsilon_{hBN}$=4.5~\cite{geick} and the reduced effective mass $\mu$=0.2~$m_0$~\cite{stierhBN} of the WSe$_2$ ML. Intriguingly,
the extracted $\delta$-parameter is close to zero which implies that the $\epsilon_n$=$E_{ns}$-$E_g$ Rydberg series found in a 2D system resembles
that of a 3D hydrogen atom ($\epsilon_n$$\sim$$-1/n^2$).

\begin{figure}[b]
    \includegraphics[width=6.5cm]{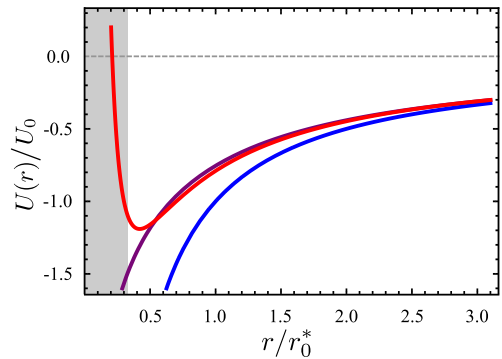}
    \caption{\label{fig:potentials}
        Rytova-Keldysh (purple curve), Coulomb (blue curve) and Kratzer potential with $g^2=0.21$ (red curve), as a function of dimensionless parameter
        $r/r_0^*$. The energy scale is measured in units of $U_0=e^2/r_0$. The grey rectangular depicts the region of distances smaller than the lattice
        constant $a=3.28\,\mbox{{\AA}}$ of WSe$_2$ ML encapsulated in hBN ($r_0^*=10\,\mbox{{\AA}}$).}
\end{figure}

On the theoretical ground, the problem of excitonic spectrum in
S-TMD MLs is commonly solved by invoking the Rytova-Keldysh
potential~\cite{rytova,keldysh} $U_{RK}(r)$ (see purple curve in
Fig.~\ref{fig:potentials}) to account for a specific character of
the $e$-$h$ attraction in these systems. At large \mbox{$e$-$h$}
distances $r$, $U_{RK}(r)$ coincides with a usual Coulomb
potential $U_{RK}(r)\sim -e^2/\varepsilon r$ (see blue curve in
Fig.\ref{fig:potentials}), which scales with the dielectric
constant $\varepsilon$ of the material surrounding the monolayer.
On the other hand, $U_{RK}(r)\sim \log(r\varepsilon/r_0)$ when $r$
is small, what accounts for the effective dielectric screening
length $r_0=2\pi \chi_{2D}$ in the system, where $\chi_{2D}$ is
the 2D polarizability of S-TMD ML. Distinctly, the apparent
excitonic spectra and the related exciton binding energies
critically depend on the efficiency of dielectric screening of the
electron-hole attraction in the medium surrounding the monolayer.

Whereas previous efforts have been largely focused on the numerical study of such problem, we show that our model provides the analytical
solution, which is in a good agreement with the experimental results discussed above. We propose to replace $U_{RK}(r)$ with the approximate
potential $U_{app}(r)$, taken in the form of piecewise function. Namely, the sub-function $U_{cor}(r)$ defines $U_{app}(r)$ at small distances $r$
(the core domain), while the external potential $U_{ext}(r)$ corresponds to $U_{app}(r)$ in the region outside of the core.

We choose the external potential in the form of the modified Kratzer potential~\cite{kratzer} (given in CGS units)
\begin{equation}
\label{eq:kratzer_potential}
U_{ext}(r)=-\frac{e^2}{r_0}\Big[\frac{r_0^*}{r}-\frac{g^2r_0^{*2}}{r^2}\Big],
\end{equation}
where $r_0^*=r_0/\varepsilon$ is the reduced screening length and $g$ is a tunable parameter.
For the case of $g^2=0.21$, $U_{ext}(r)$ fits $U_{RK}(r)$ in the region $r>r_{min}=0.46$~$r_0^*$ with the relative deviation less than $5\%$.
For the WSe$_2$ ML encapsulated in hBN, the minimal distance $r_{min}=4.6\,\mbox{{\AA}}$ is comparable with the lattice constant
$a=3.28\,\mbox{{\AA}}$~\cite{kormanyos} of WSe$_2$ (see Fig~\ref{fig:potentials} for comparison).

The Schr\"{o}dinger equation with the Kratzer potential~(\ref{eq:kratzer_potential}) is exactly solvable providing the excitonic spectrum of the
$s$-type states (see SM for details):
\begin{equation}
\label{eq:spectrum_kratzer_ns}
\epsilon_n=-Ry^*/(n+g\kappa-1/2)^2,
\end{equation}
in which $\kappa^2=2r_0^*/a_B^*$ and $a_B^*=\hbar^2\varepsilon/\mu e^2$ is the effective Bohr radius.
The effective Rydberg constant $Ry^*=e^2/2\varepsilon a_B^*$ sets the energy scale in the system, while $\delta=g\kappa-1/2$ defines the relative positions
of the energy levels in the spectrum. Since $g\kappa \propto \sqrt{\mu r_0}/\varepsilon$, the parameter $\delta$ is system dependent and its
value can be tuned, in particular, by modifying the dielectric constant $\varepsilon$. We note that for a given material, $Ry^*\propto 1/\varepsilon^2$ and
$\delta+1/2\propto 1/\varepsilon$. Such scaling laws as well as the energy sequence (\ref{eq:spectrum_kratzer_ns}) can be derived from numerical simulations
with $U_{RK}(r)$ potential at relatively large $\varepsilon$ (see SM for details). Note that Eq. \ref{eq:spectrum_kratzer_ns} is an analogous of our experimentally
found relation given by Eq.~\ref{eq:eq_1}.

In the following, we introduce $U_{cor}(r)$ which replaces the Kratzer potential at small distances $r$, comparable with the lattice constant of WSe$_2$ in our
particular case. We choose the constant attractive potential $U_{cor}(r)=V_0$. Below we demonstrate, that $U_{cor}(r)$ does not change $\propto(n+\delta)^{-2}$
behaviour of the spectrum and modifies only $\delta$ parameter.

We consider the Kratzer and constant potentials as external and core ones, respectively. We choose the parameter $g^2=0.21$ and the region of validity
of the Kratzer potential up to its minimum $\xi_0=2g^2$, where $\xi=r/r_0^*$. The parameter $V_0$ of the core potential is chosen as an average value of
$U_{RK}(\xi)$ in the domain $\xi\in[0,\xi_0]$: $V_0=2\xi_0^{-2}\int_0^{\xi_0} d\xi \xi U_{RK}(\xi)$. Finally the approximate potential is
\begin{equation}
U_{app}(\xi)=-U_0\Big\{\Big[\frac{1}{\xi}-\frac{0.21}{\xi^2}\Big]\theta(\xi-\xi_0)+v_0\,\theta(\xi_0-\xi)\Big\},
\end{equation}
where $\theta(x)$ is the step-function and $v_0=1.71134$. Please,
note that the truncated Kratzer potential is applicable only if
the radius of the core potential $r_{cor}$ is less or comparable
with the lattice constant $a$. We estimate that this requirement
is well satisfied for all monolayers encapsulated in hBN, and in
particular for our WSe$_2$ structure, for which
$r_{cor}=2g^2r_0^*=4.2\,\mbox{\AA}$ and $a=3.28\,\mbox{\AA}$.
Considering the $s$-type excitonic states in this later system, we
derive the following formula (a detailed description is given in
SM)
\begin{equation}
\epsilon_n=-134\,\text{meV}/(n-0.099)^2.
\end{equation}
Both found values: $Ry^*=134\,\text{meV}$ and $\delta=-0.099$
match their experimentally obtained counterparts (with the aid of
Eq.~\ref*{eq:eq_1}) 140.5~meV and -0.083, respectively.

The applicability range of the formula given by Eq.~\ref*{eq:eq_1}
can be also considered from a different angle, {\it i.e.}, when it is
directly compared/fitted to numerical solutions obtained within
the Rytova-Keldysh formalism. As demonstrated in the SM, the
validity range of Eq.~\ref*{eq:eq_1} can be defined with respect
to a single, dimensionless parameter of a monolayer structure:
$b=a_B^*/r_0^*$, and we find that our simple approach is valid
when $b>0.3$, and estimate that this condition is well satisfied
for all monolayers encapsulated in hBN. Nevertheless, even if $b$
is as small as $b\approx 0.1$, what may correspond to the case a
monolayer deposited on Si/SiO$_2$ substrate, the spectrum given by
Eq.~\ref*{eq:eq_1} coincides with that derived with the
Rytova-Keldysh potential within the accuracy of $5\%$.

\begin{figure}[t]
    \centering
    \includegraphics[width=1\linewidth]{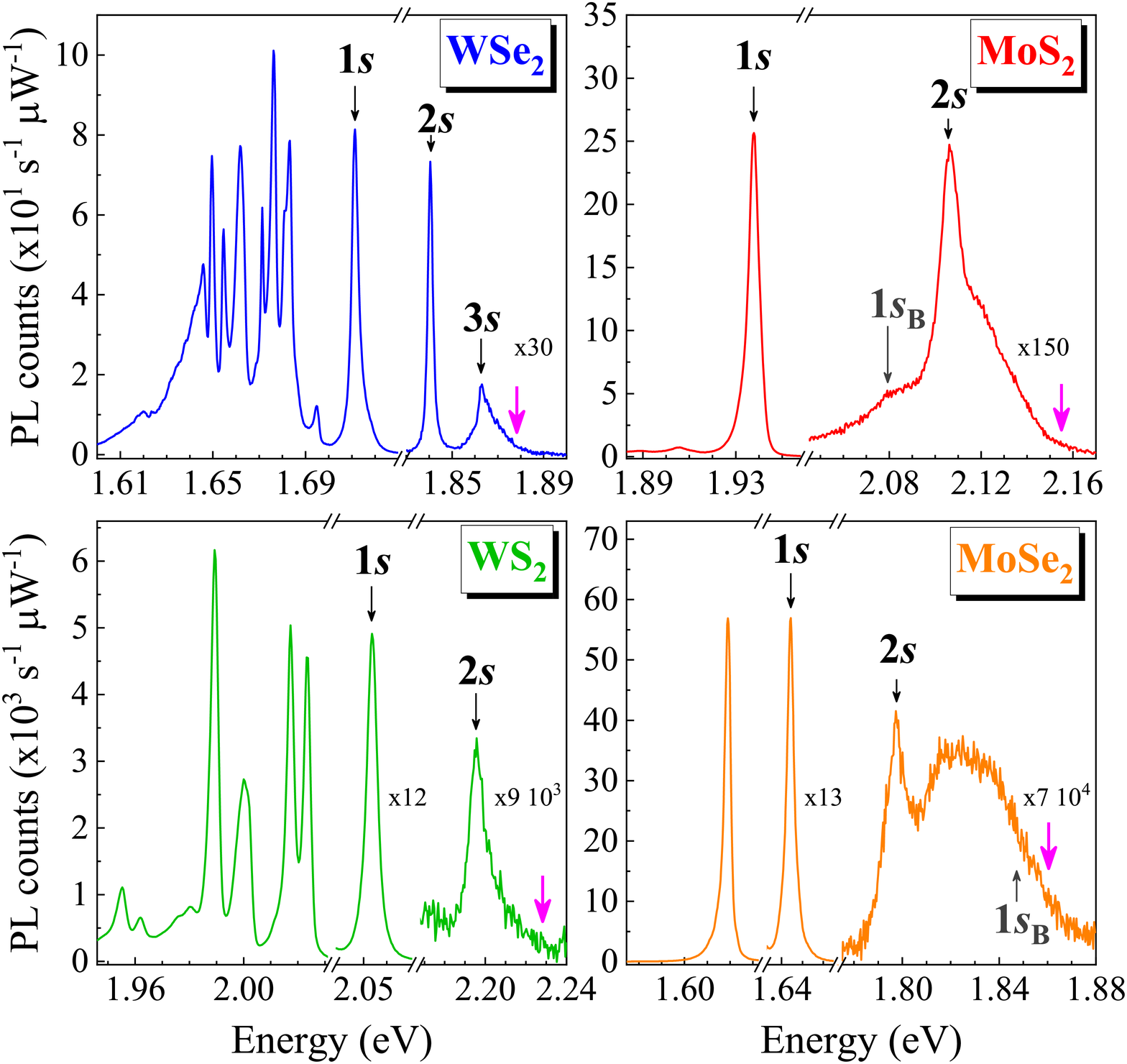}%
    \caption{Low temperature PL spectra of S-TMD MLs at $T$=5~K. The pink vertical arrows denote the estimated band-gap energies $E_g$. The chosen spectral regions are scaled for clarity. Typically for S-TMDs monolayers, the most pronounced emission feature seen in our spectra is due to the 1$s$ excitonic resonance accompanied by low energy peaks commonly assigned to different excitonic complexes~\cite{CadizPRX,wierzbowski2017,chen2017,courtade2017,Robert2017,Vaclavkova2018,nagler2018,robert2018,Barbone2018,Chen2018,Li2018,molasWSe2,liuTRIONS}.}
    \label{fig:fig_5}
\end{figure}

\begin{center}
    \begin{table*}[t]
        \centering
        \caption{Series of parameters ($E^{\text{exp}}_b$, $\Delta E^{\text{exp}}_{2s-1s}$, $\delta^{\text{exp}}$, $\mu^{\text{exp}}$) obtained from the analysis of PL spectra shown in Fig.\ref{fig:fig_5}, compared with results of DFT calculations ($\mu^{\text{DFT}}$ $(m_0)$)~\cite{kormanyos}.}
        \label{tab}
        \begin{tabular}{cccccc}
            \hline
            \multicolumn{1}{c}{Monolayer} & \multicolumn{1}{|c}{$E^{\text{exp}}_b$ (meV)} & \multicolumn{1}{|c}{$\Delta E^{\text{exp}}_{2s-1s}$ (meV)} & \multicolumn{1}{|c}{$\delta^{\text{exp}}$} &  \multicolumn{1}{|c}{$\mu^{\text{exp}}$ $(m_0)$} & \multicolumn{1}{|c }{$\mu^{\text{DFT}}$ $(m_0)$}  \\
            \hline
            \multicolumn{1}{c}{WSe$_2$} & \multicolumn{1}{|c}{167} & \multicolumn{1}{|c}{130} & \multicolumn{1}{|c }{-0.083}  & \multicolumn{1}{|c }{0.21} & \multicolumn{1}{|c }{0.16} \\
            \hline
            \multicolumn{1}{c}{MoSe$_2$}  & \multicolumn{1}{|c }{216} & \multicolumn{1}{|c }{153} & \multicolumn{1}{|c }{0.174} & \multicolumn{1}{|c }{0.44} & \multicolumn{1}{|c }{0.27} \\
            \hline
            \multicolumn{1}{c }{WS$_2$}  & \multicolumn{1}{|c }{174} & \multicolumn{1}{|c }{141} & \multicolumn{1}{|c }{-0.229} & \multicolumn{1}{|c }{0.15} & \multicolumn{1}{|c }{0.15} \\
            \hline
            \multicolumn{1}{c }{MoS$_2$}  & \multicolumn{1}{|c }{217} & \multicolumn{1}{|c }{168} & \multicolumn{1}{|c }{-0.095}  & \multicolumn{1}{|c }{0.26} & \multicolumn{1}{|c }{0.24} \\
            \hline
        \end{tabular}
    \end{table*}
\end{center}

The model proposed above accounts well for the experimental
results obtained for the WSe$_2$ monolayer and it is obviously
interesting to test this model for other S-TMD materials, as well.
Unfortunately, the observation of the rich Rydberg spectrum of
excitonic states in S-TMD MLs seems to be, so far, uniquely
reserved for WSe$_2$ MLs. Nevertheless, for all other S-TMD MLs
studied, $i.e.$, MoS$_2$, WS$_2$, and MoSe$_2$ MLs encapsulated in
hBN, we do experimentally observe the 2$s$ in addition to the 1$s$
excitonic resonance (PL and reflectance contrast spectra), see
Fig.~\ref{fig:fig_5} and Fig.~S7 in SM. The energy positions,
$E_{1s}$ and $E_{2s}$, of, correspondingly, the 1$s$ and 2$s$
resonances (of A exciton) are directly read from the data shown in
Fig.~\ref{fig:fig_5}. Of interest is the energy difference
($E_{2s}$-$E_{1s}$)=$\Delta E^{\text{exp}}_{2s-1s}$ listed in
Table~\ref{tab}, for all four MLs investigated. As shown in
Fig.~\ref{fig:fig_5}, the PL peaks associated with the excited
excitonic states are followed by noticeable PL tails developed at
higher energies. We believe that these PL tails penetrate above
the band-gap energies which are, however, not spectacularly marked
in the spectra. We note, however, that in the case of our
exemplary WSe$_2$ ML, the PL intensity at the band-gap energy
(accurately estimated from magneto-PL data and marked with a pink
arrow in Fig.~\ref{fig:fig_5}) consists of 5$\%$ of the intensity
of the 2$s$ exciton PL peak. Applying the same convention to all
spectra presented in Fig.~\ref{fig:fig_5}, we estimate the band
gaps in the three other MLs, as illustrated with pink arrows in
this figure. Most critical is estimation of the band gap in
MoSe$_2$ ML, which requires a convolution of the PL spectra due to
an additional signal associated with the B-exciton resonance (see
SM for details). With estimation of the band gap and reading the
energies of 1$s$ excitonic resonances directly from the spectra
(see Fig.~\ref{fig:fig_5}), we extract exciton binding energies
$E_b^{exp}=(E_g-E_{1s})$ and show their values in Table~\ref{tab}.
Having estimated $\Delta E^{\text{exp}}_{2s-1s}$ and $E_b^{exp}$
parameters, and following our predictions that
$E_{ns}=E_g-Ry^*/{(n+\delta)^2}$, where $Ry^*=Ry\times
\mu/(\varepsilon_{hBN}^2m_0)$, we derive the $\delta^{exp}$ and
$\mu^{exp}$ parameters for all MLs studied, see Table~\ref{tab}.
We found very good agreement between our estimations and results
of DFT calculations~\cite{kormanyos} for the reduced masses in
WS$_2$ and MoS$_2$ MLs, while we note an apparent discrepancy for
WSe$_2$ and MoSe$_2$ MLs. We also applied our
model to estimate values of band gaps and binding energies to the
experimental data available in the
literature~\cite{chernikov,goryca2019}, which is discussed in SM.

Concluding, the presented experimental and theoretical study let us proposed that the $ns$ Rydberg series of excitonic
states in S-TMD monolayers encapsulated in hBN follows a simple energy ladder: $\epsilon_n = -Ry^*  /(n+\delta)^2$. $Ry^*=Ry\times \mu/(\varepsilon^2m_0)$,
where $Ry$ is the Rydberg energy, $\mu$ denotes the reduced $e$-$h$ mass, and $\varepsilon$ is the dielectric
constant  of the surrounding material. The dielectric polarizability $\chi_{2D}$ of a monolayer is only
encoded in $\delta$ which, in the first approximation, is given by $\delta=0.21\kappa-1/2$, where
$\kappa^2=2\mu r_0e^2/\hbar^2\varepsilon^2$ and $r_0=2\pi \chi_{2D}$ is a characteristic 2D screening length.
Strikingly, $\delta$ is found  to be close to zero for WSe$_2$ (and MoS$_2$) ML whose $\epsilon_n$ spectrum
resembles that of a 3D hydrogen atom. The proposed model may be applicable to other Coulomb bound states
($e.g.$ donor and/or acceptor states), also to other systems, such as colloidal platelets~\cite{Ithurria2011}
or 2D perovskites~\cite{Stoumpos2016}. Finally, we note that our $\epsilon_n=-Ry^*/(n+\delta)^2$ solution
coincides with that expected for a hypothetical hydrogen atom in fractional dimension $N$, ($N=2\delta+3$), which
was indeed speculated~\cite{christol} to mimic the spectrum of Coulomb bound states in low-dimensional
semiconductor structures.

The work has been supported by the EU Graphene Flagship project (no. 785219), the ATOMOPTO project (TEAM programme of the Foundation for Polish Science, co-financed by the EU within the ERD-Fund), the National Science Centre, Poland (grants no. 2013/10/M/ST3/00791, 2017/27/B/ST3/00205, and 2018/31/B/ST3/02111), and the Nanofab facility of the Institut N\'eel, CNRS UGA, and the LNCMI-CNRS, a member of the European Magnetic Field Laboratory (EMFL). M.B. acknowledges the financial support from the Ministry of Education, Youth and Sports of the Czech Republic under the project CEITEC 2020 (LQ1601). K.W. and T.T. acknowledge support from the Elemental Strategy Initiative conducted by the MEXT, Japan, and the CREST (JPMJCR15F3), JST.

\bibliographystyle{apsrev4-1}
\bibliography{biblio_ES}

\begin{thebibliography}{39}%
\makeatletter
\providecommand \@ifxundefined [1]{%
 \@ifx{#1\undefined}
}%
\providecommand \@ifnum [1]{%
 \ifnum #1\expandafter \@firstoftwo
 \else \expandafter \@secondoftwo
 \fi
}%
\providecommand \@ifx [1]{%
 \ifx #1\expandafter \@firstoftwo
 \else \expandafter \@secondoftwo
 \fi
}%
\providecommand \natexlab [1]{#1}%
\providecommand \enquote  [1]{``#1''}%
\providecommand \bibnamefont  [1]{#1}%
\providecommand \bibfnamefont [1]{#1}%
\providecommand \citenamefont [1]{#1}%
\providecommand \href@noop [0]{\@secondoftwo}%
\providecommand \href [0]{\begingroup \@sanitize@url \@href}%
\providecommand \@href[1]{\@@startlink{#1}\@@href}%
\providecommand \@@href[1]{\endgroup#1\@@endlink}%
\providecommand \@sanitize@url [0]{\catcode `\\12\catcode `\$12\catcode
  `\&12\catcode `\#12\catcode `\^12\catcode `\_12\catcode `\%12\relax}%
\providecommand \@@startlink[1]{}%
\providecommand \@@endlink[0]{}%
\providecommand \url  [0]{\begingroup\@sanitize@url \@url }%
\providecommand \@url [1]{\endgroup\@href {#1}{\urlprefix }}%
\providecommand \urlprefix  [0]{URL }%
\providecommand \Eprint [0]{\href }%
\providecommand \doibase [0]{http://dx.doi.org/}%
\providecommand \selectlanguage [0]{\@gobble}%
\providecommand \bibinfo  [0]{\@secondoftwo}%
\providecommand \bibfield  [0]{\@secondoftwo}%
\providecommand \translation [1]{[#1]}%
\providecommand \BibitemOpen [0]{}%
\providecommand \bibitemStop [0]{}%
\providecommand \bibitemNoStop [0]{.\EOS\space}%
\providecommand \EOS [0]{\spacefactor3000\relax}%
\providecommand \BibitemShut  [1]{\csname bibitem#1\endcsname}%
\let\auto@bib@innerbib\@empty
\bibitem [{\citenamefont {Rytova}(1967)}]{rytova}%
  \BibitemOpen
  \bibfield  {author} {\bibinfo {author} {\bibfnamefont {N.~S.}\ \bibnamefont
  {Rytova}},\ }\href@noop {} {\bibfield  {journal} {\bibinfo  {journal} {Proc.
  MSU, Phys. Astron.}\ }\textbf {\bibinfo {volume} {3}},\ \bibinfo {pages}
  {308} (\bibinfo {year} {1967})}\BibitemShut {NoStop}%
\bibitem [{\citenamefont {Keldysh}(1979)}]{keldysh}%
  \BibitemOpen
  \bibfield  {author} {\bibinfo {author} {\bibfnamefont {L.~V.}\ \bibnamefont
  {Keldysh}},\ }\href@noop {} {\bibfield  {journal} {\bibinfo  {journal} {JETP
  Lett.}\ }\textbf {\bibinfo {volume} {29}},\ \bibinfo {pages} {658} (\bibinfo
  {year} {1979})}\BibitemShut {NoStop}%
\bibitem [{\citenamefont {Kotov}\ \emph {et~al.}(2012)\citenamefont {Kotov},
  \citenamefont {Uchoa}, \citenamefont {Pereira}, \citenamefont {Guinea},\ and\
  \citenamefont {Castro~Neto}}]{kotov2012}%
  \BibitemOpen
  \bibfield  {author} {\bibinfo {author} {\bibfnamefont {V.~N.}\ \bibnamefont
  {Kotov}}, \bibinfo {author} {\bibfnamefont {B.}~\bibnamefont {Uchoa}},
  \bibinfo {author} {\bibfnamefont {V.~M.}\ \bibnamefont {Pereira}}, \bibinfo
  {author} {\bibfnamefont {F.}~\bibnamefont {Guinea}}, \ and\ \bibinfo {author}
  {\bibfnamefont {A.~H.}\ \bibnamefont {Castro~Neto}},\ }\href {\doibase
  10.1103/RevModPhys.84.1067} {\bibfield  {journal} {\bibinfo  {journal} {Rev.
  Mod. Phys.}\ }\textbf {\bibinfo {volume} {84}},\ \bibinfo {pages} {1067}
  (\bibinfo {year} {2012})}\BibitemShut {NoStop}%
\bibitem [{\citenamefont {Faugeras}\ \emph {et~al.}(2015)\citenamefont
  {Faugeras}, \citenamefont {Berciaud}, \citenamefont {Leszczynski},
  \citenamefont {Henni}, \citenamefont {Nogajewski}, \citenamefont {Orlita},
  \citenamefont {Taniguchi}, \citenamefont {Watanabe}, \citenamefont
  {Forsythe}, \citenamefont {Kim}, \citenamefont {Jalil}, \citenamefont {Geim},
  \citenamefont {Basko},\ and\ \citenamefont {Potemski}}]{faugeras2015}%
  \BibitemOpen
  \bibfield  {author} {\bibinfo {author} {\bibfnamefont {C.}~\bibnamefont
  {Faugeras}}, \bibinfo {author} {\bibfnamefont {S.}~\bibnamefont {Berciaud}},
  \bibinfo {author} {\bibfnamefont {P.}~\bibnamefont {Leszczynski}}, \bibinfo
  {author} {\bibfnamefont {Y.}~\bibnamefont {Henni}}, \bibinfo {author}
  {\bibfnamefont {K.}~\bibnamefont {Nogajewski}}, \bibinfo {author}
  {\bibfnamefont {M.}~\bibnamefont {Orlita}}, \bibinfo {author} {\bibfnamefont
  {T.}~\bibnamefont {Taniguchi}}, \bibinfo {author} {\bibfnamefont
  {K.}~\bibnamefont {Watanabe}}, \bibinfo {author} {\bibfnamefont
  {C.}~\bibnamefont {Forsythe}}, \bibinfo {author} {\bibfnamefont
  {P.}~\bibnamefont {Kim}}, \bibinfo {author} {\bibfnamefont {R.}~\bibnamefont
  {Jalil}}, \bibinfo {author} {\bibfnamefont {A.~K.}\ \bibnamefont {Geim}},
  \bibinfo {author} {\bibfnamefont {D.~M.}\ \bibnamefont {Basko}}, \ and\
  \bibinfo {author} {\bibfnamefont {M.}~\bibnamefont {Potemski}},\ }\href
  {\doibase 10.1103/PhysRevLett.114.126804} {\bibfield  {journal} {\bibinfo
  {journal} {Phys. Rev. Lett.}\ }\textbf {\bibinfo {volume} {114}},\ \bibinfo
  {pages} {126804} (\bibinfo {year} {2015})}\BibitemShut {NoStop}%
\bibitem [{\citenamefont {Geim}\ and\ \citenamefont
  {Grigorieva}(2013)}]{Geim2013}%
  \BibitemOpen
  \bibfield  {author} {\bibinfo {author} {\bibfnamefont {A.~K.}\ \bibnamefont
  {Geim}}\ and\ \bibinfo {author} {\bibfnamefont {I.}~\bibnamefont
  {Grigorieva}},\ }\href {\doibase 10.1038/nature12385} {\bibfield  {journal}
  {\bibinfo  {journal} {Nature}\ }\textbf {\bibinfo {volume} {499}},\ \bibinfo
  {pages} {419} (\bibinfo {year} {2013})}\BibitemShut {NoStop}%
\bibitem [{\citenamefont {Ithurria}\ \emph {et~al.}(2011)\citenamefont
  {Ithurria}, \citenamefont {Tessier}, \citenamefont {Mahler}, \citenamefont
  {Lobo},\ and\ \citenamefont {Dubertret}}]{Ithurria2011}%
  \BibitemOpen
  \bibfield  {author} {\bibinfo {author} {\bibfnamefont {S.}~\bibnamefont
  {Ithurria}}, \bibinfo {author} {\bibfnamefont {M.~D.}\ \bibnamefont
  {Tessier}}, \bibinfo {author} {\bibfnamefont {B.}~\bibnamefont {Mahler}},
  \bibinfo {author} {\bibfnamefont {R.~P. S.~M.}\ \bibnamefont {Lobo}}, \ and\
  \bibinfo {author} {\bibfnamefont {A.~L.}\ \bibnamefont {Dubertret},
  \bibfnamefont {B.and~Efros}},\ }\href {\doibase 10.1038/nmat3145} {\bibfield
  {journal} {\bibinfo  {journal} {Nature Materials}\ }\textbf {\bibinfo
  {volume} {10}},\ \bibinfo {pages} {936} (\bibinfo {year} {2011})}\BibitemShut
  {NoStop}%
\bibitem [{\citenamefont {Stoumpos}\ \emph {et~al.}(2016)\citenamefont
  {Stoumpos}, \citenamefont {Cao}, \citenamefont {Clark}, \citenamefont
  {Young}, \citenamefont {Rondinelli}, \citenamefont {Jang}, \citenamefont
  {Hupp},\ and\ \citenamefont {Kanatzidis}}]{Stoumpos2016}%
  \BibitemOpen
  \bibfield  {author} {\bibinfo {author} {\bibfnamefont {C.~C.}\ \bibnamefont
  {Stoumpos}}, \bibinfo {author} {\bibfnamefont {D.~H.}\ \bibnamefont {Cao}},
  \bibinfo {author} {\bibfnamefont {D.~J.}\ \bibnamefont {Clark}}, \bibinfo
  {author} {\bibfnamefont {J.}~\bibnamefont {Young}}, \bibinfo {author}
  {\bibfnamefont {J.~M.}\ \bibnamefont {Rondinelli}}, \bibinfo {author}
  {\bibfnamefont {J.~I.}\ \bibnamefont {Jang}}, \bibinfo {author}
  {\bibfnamefont {J.~T.}\ \bibnamefont {Hupp}}, \ and\ \bibinfo {author}
  {\bibfnamefont {M.~G.}\ \bibnamefont {Kanatzidis}},\ }\href {\doibase
  10.1021/acs.chemmater.6b00847} {\bibfield  {journal} {\bibinfo  {journal}
  {Chemistry of Materials}\ }\textbf {\bibinfo {volume} {28}},\ \bibinfo
  {pages} {2852} (\bibinfo {year} {2016})}\BibitemShut {NoStop}%
\bibitem [{\citenamefont {Blancon}\ \emph {et~al.}(2018)\citenamefont
  {Blancon}, \citenamefont {Stier}, \citenamefont {Tsai}, \citenamefont
  {Stoumpos}, \citenamefont {Traor{\'e}}, \citenamefont {Pedesseau},
  \citenamefont {Kepenekian}, \citenamefont {Katsutani}, \citenamefont {Noe},
  \citenamefont {Kono}, \citenamefont {Tretiak}, \citenamefont {Crooker},
  \citenamefont {Katan}, \citenamefont {Kanatzidis}, \citenamefont {Crochet},
  \citenamefont {Even},\ and\ \citenamefont {Mohite}}]{Blancon2018}%
  \BibitemOpen
  \bibfield  {author} {\bibinfo {author} {\bibfnamefont {J.-C.}\ \bibnamefont
  {Blancon}}, \bibinfo {author} {\bibfnamefont {A.~V.}\ \bibnamefont {Stier}},
  \bibinfo {author} {\bibfnamefont {W.}~\bibnamefont {Tsai}, \bibfnamefont
  {H.and~Nie}}, \bibinfo {author} {\bibfnamefont {C.~C.}\ \bibnamefont
  {Stoumpos}}, \bibinfo {author} {\bibfnamefont {B.}~\bibnamefont
  {Traor{\'e}}}, \bibinfo {author} {\bibfnamefont {L.}~\bibnamefont
  {Pedesseau}}, \bibinfo {author} {\bibfnamefont {M.}~\bibnamefont
  {Kepenekian}}, \bibinfo {author} {\bibfnamefont {F.}~\bibnamefont
  {Katsutani}}, \bibinfo {author} {\bibfnamefont {G.~T.}\ \bibnamefont {Noe}},
  \bibinfo {author} {\bibfnamefont {J.}~\bibnamefont {Kono}}, \bibinfo {author}
  {\bibfnamefont {S.}~\bibnamefont {Tretiak}}, \bibinfo {author} {\bibfnamefont
  {S.~A.}\ \bibnamefont {Crooker}}, \bibinfo {author} {\bibfnamefont
  {C.}~\bibnamefont {Katan}}, \bibinfo {author} {\bibfnamefont {M.~G.}\
  \bibnamefont {Kanatzidis}}, \bibinfo {author} {\bibfnamefont {J.~J.}\
  \bibnamefont {Crochet}}, \bibinfo {author} {\bibfnamefont {J.}~\bibnamefont
  {Even}}, \ and\ \bibinfo {author} {\bibfnamefont {A.~D.}\ \bibnamefont
  {Mohite}},\ }\href {\doibase 10.1038/s41467-018-04659-x} {\bibfield
  {journal} {\bibinfo  {journal} {Nature Communications}\ }\textbf {\bibinfo
  {volume} {9}},\ \bibinfo {pages} {2254} (\bibinfo {year} {2018})}\BibitemShut
  {NoStop}%
\bibitem [{\citenamefont {Chernikov}\ \emph {et~al.}(2014)\citenamefont
  {Chernikov}, \citenamefont {Berkelbach}, \citenamefont {Hill}, \citenamefont
  {Rigosi}, \citenamefont {Li}, \citenamefont {Aslan}, \citenamefont
  {Reichman}, \citenamefont {Hybertsen},\ and\ \citenamefont
  {Heinz}}]{chernikov}%
  \BibitemOpen
  \bibfield  {author} {\bibinfo {author} {\bibfnamefont {A.}~\bibnamefont
  {Chernikov}}, \bibinfo {author} {\bibfnamefont {T.~C.}\ \bibnamefont
  {Berkelbach}}, \bibinfo {author} {\bibfnamefont {H.~M.}\ \bibnamefont
  {Hill}}, \bibinfo {author} {\bibfnamefont {A.}~\bibnamefont {Rigosi}},
  \bibinfo {author} {\bibfnamefont {Y.}~\bibnamefont {Li}}, \bibinfo {author}
  {\bibfnamefont {O.~B.}\ \bibnamefont {Aslan}}, \bibinfo {author}
  {\bibfnamefont {D.~R.}\ \bibnamefont {Reichman}}, \bibinfo {author}
  {\bibfnamefont {M.~S.}\ \bibnamefont {Hybertsen}}, \ and\ \bibinfo {author}
  {\bibfnamefont {T.~F.}\ \bibnamefont {Heinz}},\ }\href {\doibase
  10.1103/PhysRevLett.113.076802} {\bibfield  {journal} {\bibinfo  {journal}
  {Phys. Rev. Lett.}\ }\textbf {\bibinfo {volume} {113}},\ \bibinfo {pages}
  {076802} (\bibinfo {year} {2014})}\BibitemShut {NoStop}%
\bibitem [{\citenamefont {Raja}\ \emph {et~al.}(2017)\citenamefont {Raja},
  \citenamefont {Chaves}, \citenamefont {Yu}, \citenamefont {Arefe},
  \citenamefont {Hill}, \citenamefont {Rigosi}, \citenamefont {Berkelbach},
  \citenamefont {Nagler}, \citenamefont {Sch{\"u}ller}, \citenamefont {Korn},
  \citenamefont {Nuckolls}, \citenamefont {Hone}, \citenamefont {Brus},
  \citenamefont {Heinz}, \citenamefont {Reichman},\ and\ \citenamefont
  {Chernikov}}]{Raja2017}%
  \BibitemOpen
  \bibfield  {author} {\bibinfo {author} {\bibfnamefont {A.}~\bibnamefont
  {Raja}}, \bibinfo {author} {\bibfnamefont {A.}~\bibnamefont {Chaves}},
  \bibinfo {author} {\bibfnamefont {J.}~\bibnamefont {Yu}}, \bibinfo {author}
  {\bibfnamefont {G.}~\bibnamefont {Arefe}}, \bibinfo {author} {\bibfnamefont
  {H.~M.}\ \bibnamefont {Hill}}, \bibinfo {author} {\bibfnamefont {A.~F.}\
  \bibnamefont {Rigosi}}, \bibinfo {author} {\bibfnamefont {T.~C.}\
  \bibnamefont {Berkelbach}}, \bibinfo {author} {\bibfnamefont
  {P.}~\bibnamefont {Nagler}}, \bibinfo {author} {\bibfnamefont
  {C.}~\bibnamefont {Sch{\"u}ller}}, \bibinfo {author} {\bibfnamefont
  {T.}~\bibnamefont {Korn}}, \bibinfo {author} {\bibfnamefont {C.}~\bibnamefont
  {Nuckolls}}, \bibinfo {author} {\bibfnamefont {J.}~\bibnamefont {Hone}},
  \bibinfo {author} {\bibfnamefont {L.~E.}\ \bibnamefont {Brus}}, \bibinfo
  {author} {\bibfnamefont {T.~F.}\ \bibnamefont {Heinz}}, \bibinfo {author}
  {\bibfnamefont {D.~R.}\ \bibnamefont {Reichman}}, \ and\ \bibinfo {author}
  {\bibfnamefont {A.}~\bibnamefont {Chernikov}},\ }\href {\doibase
  10.1038/ncomms15251} {\bibfield  {journal} {\bibinfo  {journal} {Nature
  Communications}\ }\textbf {\bibinfo {volume} {8}},\ \bibinfo {pages} {15251}
  (\bibinfo {year} {2017})}\BibitemShut {NoStop}%
\bibitem [{\citenamefont {Koperski}\ \emph {et~al.}(2017)\citenamefont
  {Koperski}, \citenamefont {Molas}, \citenamefont {Arora}, \citenamefont
  {Nogajewski}, \citenamefont {Slobodeniuk}, \citenamefont {Faugeras},\ and\
  \citenamefont {Potemski}}]{koperski}%
  \BibitemOpen
  \bibfield  {author} {\bibinfo {author} {\bibfnamefont {M.}~\bibnamefont
  {Koperski}}, \bibinfo {author} {\bibfnamefont {M.~R.}\ \bibnamefont {Molas}},
  \bibinfo {author} {\bibfnamefont {A.}~\bibnamefont {Arora}}, \bibinfo
  {author} {\bibfnamefont {K.}~\bibnamefont {Nogajewski}}, \bibinfo {author}
  {\bibfnamefont {A.}~\bibnamefont {Slobodeniuk}}, \bibinfo {author}
  {\bibfnamefont {C.}~\bibnamefont {Faugeras}}, \ and\ \bibinfo {author}
  {\bibfnamefont {M.}~\bibnamefont {Potemski}},\ }\href {\doibase
  10.1515/nanoph-2016-0165} {\bibfield  {journal} {\bibinfo  {journal}
  {Nanophotonics}\ }\textbf {\bibinfo {volume} {6}},\ \bibinfo {pages} {1289}
  (\bibinfo {year} {2017})}\BibitemShut {NoStop}%
\bibitem [{\citenamefont {Wang}\ \emph {et~al.}(2018)\citenamefont {Wang},
  \citenamefont {Chernikov}, \citenamefont {Glazov}, \citenamefont {Heinz},
  \citenamefont {Marie}, \citenamefont {Amand},\ and\ \citenamefont
  {Urbaszek}}]{Wang2018}%
  \BibitemOpen
  \bibfield  {author} {\bibinfo {author} {\bibfnamefont {G.}~\bibnamefont
  {Wang}}, \bibinfo {author} {\bibfnamefont {A.}~\bibnamefont {Chernikov}},
  \bibinfo {author} {\bibfnamefont {M.~M.}\ \bibnamefont {Glazov}}, \bibinfo
  {author} {\bibfnamefont {T.~F.}\ \bibnamefont {Heinz}}, \bibinfo {author}
  {\bibfnamefont {X.}~\bibnamefont {Marie}}, \bibinfo {author} {\bibfnamefont
  {T.}~\bibnamefont {Amand}}, \ and\ \bibinfo {author} {\bibfnamefont
  {B.}~\bibnamefont {Urbaszek}},\ }\href {\doibase
  10.1103/RevModPhys.90.021001} {\bibfield  {journal} {\bibinfo  {journal}
  {Rev. Mod. Phys.}\ }\textbf {\bibinfo {volume} {90}},\ \bibinfo {pages}
  {021001} (\bibinfo {year} {2018})}\BibitemShut {NoStop}%
\bibitem [{\citenamefont {Trushin}\ \emph {et~al.}(2018)\citenamefont
  {Trushin}, \citenamefont {Goerbig},\ and\ \citenamefont
  {Belzig}}]{goerbig2018}%
  \BibitemOpen
  \bibfield  {author} {\bibinfo {author} {\bibfnamefont {M.}~\bibnamefont
  {Trushin}}, \bibinfo {author} {\bibfnamefont {M.~O.}\ \bibnamefont
  {Goerbig}}, \ and\ \bibinfo {author} {\bibfnamefont {W.}~\bibnamefont
  {Belzig}},\ }\href {\doibase 10.1103/PhysRevLett.120.187401} {\bibfield
  {journal} {\bibinfo  {journal} {Phys. Rev. Lett.}\ }\textbf {\bibinfo
  {volume} {120}},\ \bibinfo {pages} {187401} (\bibinfo {year}
  {2018})}\BibitemShut {NoStop}%
\bibitem [{\citenamefont {MacDonald}\ and\ \citenamefont
  {Ritchie}(1986)}]{macdonald}%
  \BibitemOpen
  \bibfield  {author} {\bibinfo {author} {\bibfnamefont {A.~H.}\ \bibnamefont
  {MacDonald}}\ and\ \bibinfo {author} {\bibfnamefont {D.~S.}\ \bibnamefont
  {Ritchie}},\ }\href {\doibase 10.1103/PhysRevB.33.8336} {\bibfield  {journal}
  {\bibinfo  {journal} {Phys. Rev. B}\ }\textbf {\bibinfo {volume} {33}},\
  \bibinfo {pages} {8336} (\bibinfo {year} {1986})}\BibitemShut {NoStop}%
\bibitem [{\citenamefont {Koteles}\ and\ \citenamefont {Chi}(1988)}]{koteles}%
  \BibitemOpen
  \bibfield  {author} {\bibinfo {author} {\bibfnamefont {E.~S.}\ \bibnamefont
  {Koteles}}\ and\ \bibinfo {author} {\bibfnamefont {J.~Y.}\ \bibnamefont
  {Chi}},\ }\href {\doibase 10.1103/PhysRevB.37.6332} {\bibfield  {journal}
  {\bibinfo  {journal} {Phys. Rev. B}\ }\textbf {\bibinfo {volume} {37}},\
  \bibinfo {pages} {6332} (\bibinfo {year} {1988})}\BibitemShut {NoStop}%
\bibitem [{\citenamefont {Christol}\ \emph {et~al.}(1993)\citenamefont
  {Christol}, \citenamefont {Lefebvre},\ and\ \citenamefont
  {Mathieu}}]{christol}%
  \BibitemOpen
  \bibfield  {author} {\bibinfo {author} {\bibfnamefont {P.}~\bibnamefont
  {Christol}}, \bibinfo {author} {\bibfnamefont {P.}~\bibnamefont {Lefebvre}},
  \ and\ \bibinfo {author} {\bibfnamefont {H.}~\bibnamefont {Mathieu}},\ }\href
  {\doibase 10.1063/1.354224} {\bibfield  {journal} {\bibinfo  {journal}
  {Journal of Applied Physics}\ }\textbf {\bibinfo {volume} {74}},\ \bibinfo
  {pages} {5626} (\bibinfo {year} {1993})}\BibitemShut {NoStop}%
\bibitem [{\citenamefont {Stier}\ \emph {et~al.}(2018)\citenamefont {Stier},
  \citenamefont {Wilson}, \citenamefont {Velizhanin}, \citenamefont {Kono},
  \citenamefont {Xu},\ and\ \citenamefont {Crooker}}]{stierhBN}%
  \BibitemOpen
  \bibfield  {author} {\bibinfo {author} {\bibfnamefont {A.~V.}\ \bibnamefont
  {Stier}}, \bibinfo {author} {\bibfnamefont {N.~P.}\ \bibnamefont {Wilson}},
  \bibinfo {author} {\bibfnamefont {K.~A.}\ \bibnamefont {Velizhanin}},
  \bibinfo {author} {\bibfnamefont {J.}~\bibnamefont {Kono}}, \bibinfo {author}
  {\bibfnamefont {X.}~\bibnamefont {Xu}}, \ and\ \bibinfo {author}
  {\bibfnamefont {S.~A.}\ \bibnamefont {Crooker}},\ }\href {\doibase
  10.1103/PhysRevLett.120.057405} {\bibfield  {journal} {\bibinfo  {journal}
  {Phys. Rev. Lett.}\ }\textbf {\bibinfo {volume} {120}},\ \bibinfo {pages}
  {057405} (\bibinfo {year} {2018})}\BibitemShut {NoStop}%
\bibitem [{\citenamefont {Kratzer}(1920)}]{kratzer}%
  \BibitemOpen
  \bibfield  {author} {\bibinfo {author} {\bibfnamefont {A.}~\bibnamefont
  {Kratzer}},\ }\href {\doibase 10.1007/BF01327754} {\bibfield  {journal}
  {\bibinfo  {journal} {Zeitschrift f{\"u}r Physik}\ }\textbf {\bibinfo
  {volume} {3}},\ \bibinfo {pages} {289} (\bibinfo {year} {1920})}\BibitemShut
  {NoStop}%
\bibitem [{\citenamefont {Akimoto}\ and\ \citenamefont
  {Hasegawa}(1967)}]{akimoto}%
  \BibitemOpen
  \bibfield  {author} {\bibinfo {author} {\bibfnamefont {O.}~\bibnamefont
  {Akimoto}}\ and\ \bibinfo {author} {\bibfnamefont {H.}~\bibnamefont
  {Hasegawa}},\ }\href {\doibase 10.1143/JPSJ.22.181} {\bibfield  {journal}
  {\bibinfo  {journal} {Journal of the Physical Society of Japan}\ }\textbf
  {\bibinfo {volume} {22}},\ \bibinfo {pages} {181} (\bibinfo {year}
  {1967})}\BibitemShut {NoStop}%
\bibitem [{\citenamefont {Potemski}\ \emph {et~al.}(1991)\citenamefont
  {Potemski}, \citenamefont {Vi\~na}, \citenamefont {Bauer}, \citenamefont
  {Maan}, \citenamefont {Ploog},\ and\ \citenamefont {Weimann}}]{potemski1991}%
  \BibitemOpen
  \bibfield  {author} {\bibinfo {author} {\bibfnamefont {M.}~\bibnamefont
  {Potemski}}, \bibinfo {author} {\bibfnamefont {L.}~\bibnamefont {Vi\~na}},
  \bibinfo {author} {\bibfnamefont {G.~E.~W.}\ \bibnamefont {Bauer}}, \bibinfo
  {author} {\bibfnamefont {J.~C.}\ \bibnamefont {Maan}}, \bibinfo {author}
  {\bibfnamefont {K.}~\bibnamefont {Ploog}}, \ and\ \bibinfo {author}
  {\bibfnamefont {G.}~\bibnamefont {Weimann}},\ }\href {\doibase
  10.1103/PhysRevB.43.14707} {\bibfield  {journal} {\bibinfo  {journal} {Phys.
  Rev. B}\ }\textbf {\bibinfo {volume} {43}},\ \bibinfo {pages} {14707}
  (\bibinfo {year} {1991})}\BibitemShut {NoStop}%
\bibitem [{\citenamefont {Manca}\ \emph {et~al.}(2017)\citenamefont {Manca},
  \citenamefont {Glazov}, \citenamefont {Robert}, \citenamefont {Cadiz},
  \citenamefont {Taniguchi}, \citenamefont {Watanabe}, \citenamefont
  {Courtade}, \citenamefont {Amand}, \citenamefont {Renucci}, \citenamefont
  {Marie}, \citenamefont {Wang},\ and\ \citenamefont {Urbaszek}}]{manca2017}%
  \BibitemOpen
  \bibfield  {author} {\bibinfo {author} {\bibfnamefont {M.}~\bibnamefont
  {Manca}}, \bibinfo {author} {\bibfnamefont {M.~M.}\ \bibnamefont {Glazov}},
  \bibinfo {author} {\bibfnamefont {C.}~\bibnamefont {Robert}}, \bibinfo
  {author} {\bibfnamefont {F.}~\bibnamefont {Cadiz}}, \bibinfo {author}
  {\bibfnamefont {T.}~\bibnamefont {Taniguchi}}, \bibinfo {author}
  {\bibfnamefont {K.}~\bibnamefont {Watanabe}}, \bibinfo {author}
  {\bibfnamefont {E.}~\bibnamefont {Courtade}}, \bibinfo {author}
  {\bibfnamefont {T.}~\bibnamefont {Amand}}, \bibinfo {author} {\bibfnamefont
  {P.}~\bibnamefont {Renucci}}, \bibinfo {author} {\bibfnamefont
  {X.}~\bibnamefont {Marie}}, \bibinfo {author} {\bibfnamefont
  {G.}~\bibnamefont {Wang}}, \ and\ \bibinfo {author} {\bibfnamefont
  {B.}~\bibnamefont {Urbaszek}},\ }\href {\doibase 10.1038/ncomms14927}
  {\bibfield  {journal} {\bibinfo  {journal} {Nature Communications}\ }\textbf
  {\bibinfo {volume} {8}},\ \bibinfo {pages} {14927} (\bibinfo {year}
  {2017})}\BibitemShut {NoStop}%
\bibitem [{\citenamefont {Chow}\ \emph {et~al.}(2017)\citenamefont {Chow},
  \citenamefont {Yu}, \citenamefont {Jones}, \citenamefont {Yan}, \citenamefont
  {Mandrus}, \citenamefont {Taniguchi}, \citenamefont {Watanabe}, \citenamefont
  {Yao},\ and\ \citenamefont {Xu}}]{Chow2017}%
  \BibitemOpen
  \bibfield  {author} {\bibinfo {author} {\bibfnamefont {C.~M.}\ \bibnamefont
  {Chow}}, \bibinfo {author} {\bibfnamefont {H.}~\bibnamefont {Yu}}, \bibinfo
  {author} {\bibfnamefont {A.~M.}\ \bibnamefont {Jones}}, \bibinfo {author}
  {\bibfnamefont {J.}~\bibnamefont {Yan}}, \bibinfo {author} {\bibfnamefont
  {D.~G.}\ \bibnamefont {Mandrus}}, \bibinfo {author} {\bibfnamefont
  {T.}~\bibnamefont {Taniguchi}}, \bibinfo {author} {\bibfnamefont
  {K.}~\bibnamefont {Watanabe}}, \bibinfo {author} {\bibfnamefont
  {W.}~\bibnamefont {Yao}}, \ and\ \bibinfo {author} {\bibfnamefont
  {X.}~\bibnamefont {Xu}},\ }\href {\doibase 10.1021/acs.nanolett.6b04944}
  {\bibfield  {journal} {\bibinfo  {journal} {Nano Letters}\ }\textbf {\bibinfo
  {volume} {17}},\ \bibinfo {pages} {1194} (\bibinfo {year}
  {2017})}\BibitemShut {NoStop}%
\bibitem [{\citenamefont {Chen}\ \emph
  {et~al.}(2018{\natexlab{a}})\citenamefont {Chen}, \citenamefont {Goldstein},
  \citenamefont {Tong}, \citenamefont {Taniguchi}, \citenamefont {Watanabe},\
  and\ \citenamefont {Yan}}]{chen2017}%
  \BibitemOpen
  \bibfield  {author} {\bibinfo {author} {\bibfnamefont {S.-Y.}\ \bibnamefont
  {Chen}}, \bibinfo {author} {\bibfnamefont {T.}~\bibnamefont {Goldstein}},
  \bibinfo {author} {\bibfnamefont {J.}~\bibnamefont {Tong}}, \bibinfo {author}
  {\bibfnamefont {T.}~\bibnamefont {Taniguchi}}, \bibinfo {author}
  {\bibfnamefont {K.}~\bibnamefont {Watanabe}}, \ and\ \bibinfo {author}
  {\bibfnamefont {J.}~\bibnamefont {Yan}},\ }\href {\doibase
  10.1103/PhysRevLett.120.046402} {\bibfield  {journal} {\bibinfo  {journal}
  {Phys. Rev. Lett.}\ }\textbf {\bibinfo {volume} {120}},\ \bibinfo {pages}
  {046402} (\bibinfo {year} {2018}{\natexlab{a}})}\BibitemShut {NoStop}%
\bibitem [{\citenamefont {Liu}\ \emph {et~al.}(2019)\citenamefont {Liu},
  \citenamefont {van Baren}, \citenamefont {Taniguchi}, \citenamefont
  {Watanabe}, \citenamefont {Chang},\ and\ \citenamefont {Lui}}]{Liu2019}%
  \BibitemOpen
  \bibfield  {author} {\bibinfo {author} {\bibfnamefont {E.}~\bibnamefont
  {Liu}}, \bibinfo {author} {\bibfnamefont {J.}~\bibnamefont {van Baren}},
  \bibinfo {author} {\bibfnamefont {T.}~\bibnamefont {Taniguchi}}, \bibinfo
  {author} {\bibfnamefont {K.}~\bibnamefont {Watanabe}}, \bibinfo {author}
  {\bibfnamefont {Y.-C.}\ \bibnamefont {Chang}}, \ and\ \bibinfo {author}
  {\bibfnamefont {C.~H.}\ \bibnamefont {Lui}},\ }\href {\doibase
  10.1103/PhysRevB.99.205420} {\bibfield  {journal} {\bibinfo  {journal} {Phys.
  Rev. B}\ }\textbf {\bibinfo {volume} {99}},\ \bibinfo {pages} {205420}
  (\bibinfo {year} {2019})}\BibitemShut {NoStop}%
\bibitem [{\citenamefont {Geick}\ \emph {et~al.}(1966)\citenamefont {Geick},
  \citenamefont {Perry},\ and\ \citenamefont {Rupprecht}}]{geick}%
  \BibitemOpen
  \bibfield  {author} {\bibinfo {author} {\bibfnamefont {R.}~\bibnamefont
  {Geick}}, \bibinfo {author} {\bibfnamefont {C.~H.}\ \bibnamefont {Perry}}, \
  and\ \bibinfo {author} {\bibfnamefont {G.}~\bibnamefont {Rupprecht}},\ }\href
  {\doibase 10.1103/PhysRev.146.543} {\bibfield  {journal} {\bibinfo  {journal}
  {Phys. Rev.}\ }\textbf {\bibinfo {volume} {146}},\ \bibinfo {pages} {543}
  (\bibinfo {year} {1966})}\BibitemShut {NoStop}%
\bibitem [{\citenamefont {Korm\'anyos}\ \emph {et~al.}(2015)\citenamefont
  {Korm\'anyos}, \citenamefont {Burkard}, \citenamefont {Gmitra}, \citenamefont
  {Fabian}, \citenamefont {Z\'olyomi}, \citenamefont {Drummond},\ and\
  \citenamefont {Fal'ko}}]{kormanyos}%
  \BibitemOpen
  \bibfield  {author} {\bibinfo {author} {\bibfnamefont {A.}~\bibnamefont
  {Korm\'anyos}}, \bibinfo {author} {\bibfnamefont {G.}~\bibnamefont
  {Burkard}}, \bibinfo {author} {\bibfnamefont {M.}~\bibnamefont {Gmitra}},
  \bibinfo {author} {\bibfnamefont {J.}~\bibnamefont {Fabian}}, \bibinfo
  {author} {\bibfnamefont {V.}~\bibnamefont {Z\'olyomi}}, \bibinfo {author}
  {\bibfnamefont {N.~D.}\ \bibnamefont {Drummond}}, \ and\ \bibinfo {author}
  {\bibfnamefont {V.}~\bibnamefont {Fal'ko}},\ }\href {\doibase
  10.1088/2053-1583/2/2/022001} {\bibfield  {journal} {\bibinfo  {journal} {2D
  Materials}\ }\textbf {\bibinfo {volume} {2}},\ \bibinfo {pages} {022001}
  (\bibinfo {year} {2015})}\BibitemShut {NoStop}%
\bibitem [{\citenamefont {Cadiz}\ \emph {et~al.}(2017)\citenamefont {Cadiz},
  \citenamefont {Courtade}, \citenamefont {Robert}, \citenamefont {Wang},
  \citenamefont {Shen}, \citenamefont {Cai}, \citenamefont {Taniguchi},
  \citenamefont {Watanabe}, \citenamefont {Carrere}, \citenamefont {Lagarde},
  \citenamefont {Manca}, \citenamefont {Amand}, \citenamefont {Renucci},
  \citenamefont {Tongay}, \citenamefont {Marie},\ and\ \citenamefont
  {Urbaszek}}]{CadizPRX}%
  \BibitemOpen
  \bibfield  {author} {\bibinfo {author} {\bibfnamefont {F.}~\bibnamefont
  {Cadiz}}, \bibinfo {author} {\bibfnamefont {E.}~\bibnamefont {Courtade}},
  \bibinfo {author} {\bibfnamefont {C.}~\bibnamefont {Robert}}, \bibinfo
  {author} {\bibfnamefont {G.}~\bibnamefont {Wang}}, \bibinfo {author}
  {\bibfnamefont {Y.}~\bibnamefont {Shen}}, \bibinfo {author} {\bibfnamefont
  {H.}~\bibnamefont {Cai}}, \bibinfo {author} {\bibfnamefont {T.}~\bibnamefont
  {Taniguchi}}, \bibinfo {author} {\bibfnamefont {K.}~\bibnamefont {Watanabe}},
  \bibinfo {author} {\bibfnamefont {H.}~\bibnamefont {Carrere}}, \bibinfo
  {author} {\bibfnamefont {D.}~\bibnamefont {Lagarde}}, \bibinfo {author}
  {\bibfnamefont {M.}~\bibnamefont {Manca}}, \bibinfo {author} {\bibfnamefont
  {T.}~\bibnamefont {Amand}}, \bibinfo {author} {\bibfnamefont
  {P.}~\bibnamefont {Renucci}}, \bibinfo {author} {\bibfnamefont
  {S.}~\bibnamefont {Tongay}}, \bibinfo {author} {\bibfnamefont
  {X.}~\bibnamefont {Marie}}, \ and\ \bibinfo {author} {\bibfnamefont
  {B.}~\bibnamefont {Urbaszek}},\ }\href {\doibase 10.1103/PhysRevX.7.021026}
  {\bibfield  {journal} {\bibinfo  {journal} {Phys. Rev. X}\ }\textbf {\bibinfo
  {volume} {7}},\ \bibinfo {pages} {021026} (\bibinfo {year}
  {2017})}\BibitemShut {NoStop}%
\bibitem [{\citenamefont {Wierzbowski}\ \emph {et~al.}(2017)\citenamefont
  {Wierzbowski}, \citenamefont {Klein}, \citenamefont {Sigger}, \citenamefont
  {Straubinger}, \citenamefont {Kremser}, \citenamefont {Taniguchi},
  \citenamefont {Watanabe}, \citenamefont {Wurstbauer}, \citenamefont
  {Holleitner}, \citenamefont {Kaniber}, \citenamefont {M{\"u}ller},\ and\
  \citenamefont {Finley}}]{wierzbowski2017}%
  \BibitemOpen
  \bibfield  {author} {\bibinfo {author} {\bibfnamefont {J.}~\bibnamefont
  {Wierzbowski}}, \bibinfo {author} {\bibfnamefont {J.}~\bibnamefont {Klein}},
  \bibinfo {author} {\bibfnamefont {F.}~\bibnamefont {Sigger}}, \bibinfo
  {author} {\bibfnamefont {C.}~\bibnamefont {Straubinger}}, \bibinfo {author}
  {\bibfnamefont {M.}~\bibnamefont {Kremser}}, \bibinfo {author} {\bibfnamefont
  {T.}~\bibnamefont {Taniguchi}}, \bibinfo {author} {\bibfnamefont
  {K.}~\bibnamefont {Watanabe}}, \bibinfo {author} {\bibfnamefont
  {U.}~\bibnamefont {Wurstbauer}}, \bibinfo {author} {\bibfnamefont {A.~W.}\
  \bibnamefont {Holleitner}}, \bibinfo {author} {\bibfnamefont
  {M.}~\bibnamefont {Kaniber}}, \bibinfo {author} {\bibfnamefont
  {K.}~\bibnamefont {M{\"u}ller}}, \ and\ \bibinfo {author} {\bibfnamefont
  {J.~J.}\ \bibnamefont {Finley}},\ }\href {\doibase
  10.1038/s41598-017-09739-4} {\bibfield  {journal} {\bibinfo  {journal}
  {Scientific Reports}\ }\textbf {\bibinfo {volume} {7}},\ \bibinfo {pages}
  {12383} (\bibinfo {year} {2017})}\BibitemShut {NoStop}%
\bibitem [{\citenamefont {Courtade}\ \emph {et~al.}(2017)\citenamefont
  {Courtade}, \citenamefont {Semina}, \citenamefont {Manca}, \citenamefont
  {Glazov}, \citenamefont {Robert}, \citenamefont {Cadiz}, \citenamefont
  {Wang}, \citenamefont {Taniguchi}, \citenamefont {Watanabe}, \citenamefont
  {Pierre}, \citenamefont {Escoffier}, \citenamefont {Ivchenko}, \citenamefont
  {Renucci}, \citenamefont {Marie}, \citenamefont {Amand},\ and\ \citenamefont
  {Urbaszek}}]{courtade2017}%
  \BibitemOpen
  \bibfield  {author} {\bibinfo {author} {\bibfnamefont {E.}~\bibnamefont
  {Courtade}}, \bibinfo {author} {\bibfnamefont {M.}~\bibnamefont {Semina}},
  \bibinfo {author} {\bibfnamefont {M.}~\bibnamefont {Manca}}, \bibinfo
  {author} {\bibfnamefont {M.~M.}\ \bibnamefont {Glazov}}, \bibinfo {author}
  {\bibfnamefont {C.}~\bibnamefont {Robert}}, \bibinfo {author} {\bibfnamefont
  {F.}~\bibnamefont {Cadiz}}, \bibinfo {author} {\bibfnamefont
  {G.}~\bibnamefont {Wang}}, \bibinfo {author} {\bibfnamefont {T.}~\bibnamefont
  {Taniguchi}}, \bibinfo {author} {\bibfnamefont {K.}~\bibnamefont {Watanabe}},
  \bibinfo {author} {\bibfnamefont {M.}~\bibnamefont {Pierre}}, \bibinfo
  {author} {\bibfnamefont {W.}~\bibnamefont {Escoffier}}, \bibinfo {author}
  {\bibfnamefont {E.~L.}\ \bibnamefont {Ivchenko}}, \bibinfo {author}
  {\bibfnamefont {P.}~\bibnamefont {Renucci}}, \bibinfo {author} {\bibfnamefont
  {X.}~\bibnamefont {Marie}}, \bibinfo {author} {\bibfnamefont
  {T.}~\bibnamefont {Amand}}, \ and\ \bibinfo {author} {\bibfnamefont
  {B.}~\bibnamefont {Urbaszek}},\ }\href {\doibase 10.1103/PhysRevB.96.085302}
  {\bibfield  {journal} {\bibinfo  {journal} {Phys. Rev. B}\ }\textbf {\bibinfo
  {volume} {96}},\ \bibinfo {pages} {085302} (\bibinfo {year}
  {2017})}\BibitemShut {NoStop}%
\bibitem [{\citenamefont {Robert}\ \emph {et~al.}(2017)\citenamefont {Robert},
  \citenamefont {Amand}, \citenamefont {Cadiz}, \citenamefont {Lagarde},
  \citenamefont {Courtade}, \citenamefont {Manca}, \citenamefont {Taniguchi},
  \citenamefont {Watanabe}, \citenamefont {Urbaszek},\ and\ \citenamefont
  {Marie}}]{Robert2017}%
  \BibitemOpen
  \bibfield  {author} {\bibinfo {author} {\bibfnamefont {C.}~\bibnamefont
  {Robert}}, \bibinfo {author} {\bibfnamefont {T.}~\bibnamefont {Amand}},
  \bibinfo {author} {\bibfnamefont {F.}~\bibnamefont {Cadiz}}, \bibinfo
  {author} {\bibfnamefont {D.}~\bibnamefont {Lagarde}}, \bibinfo {author}
  {\bibfnamefont {E.}~\bibnamefont {Courtade}}, \bibinfo {author}
  {\bibfnamefont {M.}~\bibnamefont {Manca}}, \bibinfo {author} {\bibfnamefont
  {T.}~\bibnamefont {Taniguchi}}, \bibinfo {author} {\bibfnamefont
  {K.}~\bibnamefont {Watanabe}}, \bibinfo {author} {\bibfnamefont
  {B.}~\bibnamefont {Urbaszek}}, \ and\ \bibinfo {author} {\bibfnamefont
  {X.}~\bibnamefont {Marie}},\ }\href {\doibase 10.1103/PhysRevB.96.155423}
  {\bibfield  {journal} {\bibinfo  {journal} {Phys. Rev. B}\ }\textbf {\bibinfo
  {volume} {96}},\ \bibinfo {pages} {155423} (\bibinfo {year}
  {2017})}\BibitemShut {NoStop}%
\bibitem [{\citenamefont {Vaclavkova}\ \emph {et~al.}(2018)\citenamefont
  {Vaclavkova}, \citenamefont {Wyzula}, \citenamefont {Nogajewski},
  \citenamefont {Bartos}, \citenamefont {Slobodeniuk}, \citenamefont
  {Faugeras}, \citenamefont {Potemski},\ and\ \citenamefont
  {Molas}}]{Vaclavkova2018}%
  \BibitemOpen
  \bibfield  {author} {\bibinfo {author} {\bibfnamefont {D.}~\bibnamefont
  {Vaclavkova}}, \bibinfo {author} {\bibfnamefont {J.}~\bibnamefont {Wyzula}},
  \bibinfo {author} {\bibfnamefont {K.}~\bibnamefont {Nogajewski}}, \bibinfo
  {author} {\bibfnamefont {M.}~\bibnamefont {Bartos}}, \bibinfo {author}
  {\bibfnamefont {A.~O.}\ \bibnamefont {Slobodeniuk}}, \bibinfo {author}
  {\bibfnamefont {C.}~\bibnamefont {Faugeras}}, \bibinfo {author}
  {\bibfnamefont {M.}~\bibnamefont {Potemski}}, \ and\ \bibinfo {author}
  {\bibfnamefont {M.~R.}\ \bibnamefont {Molas}},\ }\href {\doibase
  10.1088/1361-6528/aac65c} {\bibfield  {journal} {\bibinfo  {journal}
  {Nanotechnology}\ }\textbf {\bibinfo {volume} {29}},\ \bibinfo {pages}
  {325705} (\bibinfo {year} {2018})}\BibitemShut {NoStop}%
\bibitem [{\citenamefont {Nagler}\ \emph {et~al.}(2018)\citenamefont {Nagler},
  \citenamefont {Ballottin}, \citenamefont {Mitioglu}, \citenamefont {Durnev},
  \citenamefont {Taniguchi}, \citenamefont {Watanabe}, \citenamefont
  {Chernikov}, \citenamefont {Sch\"uller}, \citenamefont {Glazov},
  \citenamefont {Christianen},\ and\ \citenamefont {Korn}}]{nagler2018}%
  \BibitemOpen
  \bibfield  {author} {\bibinfo {author} {\bibfnamefont {P.}~\bibnamefont
  {Nagler}}, \bibinfo {author} {\bibfnamefont {M.~V.}\ \bibnamefont
  {Ballottin}}, \bibinfo {author} {\bibfnamefont {A.~A.}\ \bibnamefont
  {Mitioglu}}, \bibinfo {author} {\bibfnamefont {M.~V.}\ \bibnamefont
  {Durnev}}, \bibinfo {author} {\bibfnamefont {T.}~\bibnamefont {Taniguchi}},
  \bibinfo {author} {\bibfnamefont {K.}~\bibnamefont {Watanabe}}, \bibinfo
  {author} {\bibfnamefont {A.}~\bibnamefont {Chernikov}}, \bibinfo {author}
  {\bibfnamefont {C.}~\bibnamefont {Sch\"uller}}, \bibinfo {author}
  {\bibfnamefont {M.~M.}\ \bibnamefont {Glazov}}, \bibinfo {author}
  {\bibfnamefont {P.~C.~M.}\ \bibnamefont {Christianen}}, \ and\ \bibinfo
  {author} {\bibfnamefont {T.}~\bibnamefont {Korn}},\ }\href {\doibase
  10.1103/PhysRevLett.121.057402} {\bibfield  {journal} {\bibinfo  {journal}
  {Phys. Rev. Lett.}\ }\textbf {\bibinfo {volume} {121}},\ \bibinfo {pages}
  {057402} (\bibinfo {year} {2018})}\BibitemShut {NoStop}%
\bibitem [{\citenamefont {Robert}\ \emph {et~al.}(2018)\citenamefont {Robert},
  \citenamefont {Semina}, \citenamefont {Cadiz}, \citenamefont {Manca},
  \citenamefont {Courtade}, \citenamefont {Taniguchi}, \citenamefont
  {Watanabe}, \citenamefont {Cai}, \citenamefont {Tongay}, \citenamefont
  {Lassagne}, \citenamefont {Renucci}, \citenamefont {Amand}, \citenamefont
  {Marie}, \citenamefont {Glazov},\ and\ \citenamefont
  {Urbaszek}}]{robert2018}%
  \BibitemOpen
  \bibfield  {author} {\bibinfo {author} {\bibfnamefont {C.}~\bibnamefont
  {Robert}}, \bibinfo {author} {\bibfnamefont {M.~A.}\ \bibnamefont {Semina}},
  \bibinfo {author} {\bibfnamefont {F.}~\bibnamefont {Cadiz}}, \bibinfo
  {author} {\bibfnamefont {M.}~\bibnamefont {Manca}}, \bibinfo {author}
  {\bibfnamefont {E.}~\bibnamefont {Courtade}}, \bibinfo {author}
  {\bibfnamefont {T.}~\bibnamefont {Taniguchi}}, \bibinfo {author}
  {\bibfnamefont {K.}~\bibnamefont {Watanabe}}, \bibinfo {author}
  {\bibfnamefont {H.}~\bibnamefont {Cai}}, \bibinfo {author} {\bibfnamefont
  {S.}~\bibnamefont {Tongay}}, \bibinfo {author} {\bibfnamefont
  {B.}~\bibnamefont {Lassagne}}, \bibinfo {author} {\bibfnamefont
  {P.}~\bibnamefont {Renucci}}, \bibinfo {author} {\bibfnamefont
  {T.}~\bibnamefont {Amand}}, \bibinfo {author} {\bibfnamefont
  {X.}~\bibnamefont {Marie}}, \bibinfo {author} {\bibfnamefont {M.~M.}\
  \bibnamefont {Glazov}}, \ and\ \bibinfo {author} {\bibfnamefont
  {B.}~\bibnamefont {Urbaszek}},\ }\href {\doibase
  10.1103/PhysRevMaterials.2.011001} {\bibfield  {journal} {\bibinfo  {journal}
  {Phys. Rev. Materials}\ }\textbf {\bibinfo {volume} {2}},\ \bibinfo {pages}
  {011001} (\bibinfo {year} {2018})}\BibitemShut {NoStop}%
\bibitem [{\citenamefont {Barbone}\ \emph {et~al.}(2018)\citenamefont
  {Barbone}, \citenamefont {Montblanch}, \citenamefont {Kara}, \citenamefont
  {Palacios-Berraquero}, \citenamefont {Cadore}, \citenamefont {De~Fazio},
  \citenamefont {Pingault}, \citenamefont {Mostaani}, \citenamefont {Li},
  \citenamefont {Chen}, \citenamefont {Watanabe}, \citenamefont {Taniguchi},
  \citenamefont {Tongay}, \citenamefont {Wang}, \citenamefont {Ferrari},\ and\
  \citenamefont {Atat{\"u}re}}]{Barbone2018}%
  \BibitemOpen
  \bibfield  {author} {\bibinfo {author} {\bibfnamefont {M.}~\bibnamefont
  {Barbone}}, \bibinfo {author} {\bibfnamefont {A.~R.~P.}\ \bibnamefont
  {Montblanch}}, \bibinfo {author} {\bibfnamefont {D.~M.}\ \bibnamefont
  {Kara}}, \bibinfo {author} {\bibfnamefont {C.}~\bibnamefont
  {Palacios-Berraquero}}, \bibinfo {author} {\bibfnamefont {A.~R.}\
  \bibnamefont {Cadore}}, \bibinfo {author} {\bibfnamefont {D.}~\bibnamefont
  {De~Fazio}}, \bibinfo {author} {\bibfnamefont {B.}~\bibnamefont {Pingault}},
  \bibinfo {author} {\bibfnamefont {E.}~\bibnamefont {Mostaani}}, \bibinfo
  {author} {\bibfnamefont {H.}~\bibnamefont {Li}}, \bibinfo {author}
  {\bibfnamefont {B.}~\bibnamefont {Chen}}, \bibinfo {author} {\bibfnamefont
  {K.}~\bibnamefont {Watanabe}}, \bibinfo {author} {\bibfnamefont
  {T.}~\bibnamefont {Taniguchi}}, \bibinfo {author} {\bibfnamefont
  {S.}~\bibnamefont {Tongay}}, \bibinfo {author} {\bibfnamefont
  {G.}~\bibnamefont {Wang}}, \bibinfo {author} {\bibfnamefont {A.~C.}\
  \bibnamefont {Ferrari}}, \ and\ \bibinfo {author} {\bibfnamefont
  {M.}~\bibnamefont {Atat{\"u}re}},\ }\href {\doibase
  10.1038/s41467-018-05632-4} {\bibfield  {journal} {\bibinfo  {journal}
  {Nature Communications}\ }\textbf {\bibinfo {volume} {9}},\ \bibinfo {pages}
  {3721} (\bibinfo {year} {2018})}\BibitemShut {NoStop}%
\bibitem [{\citenamefont {Chen}\ \emph
  {et~al.}(2018{\natexlab{b}})\citenamefont {Chen}, \citenamefont {Goldstein},
  \citenamefont {Taniguchi}, \citenamefont {Watanabe},\ and\ \citenamefont
  {Yan}}]{Chen2018}%
  \BibitemOpen
  \bibfield  {author} {\bibinfo {author} {\bibfnamefont {S.-Y.}\ \bibnamefont
  {Chen}}, \bibinfo {author} {\bibfnamefont {T.}~\bibnamefont {Goldstein}},
  \bibinfo {author} {\bibfnamefont {T.}~\bibnamefont {Taniguchi}}, \bibinfo
  {author} {\bibfnamefont {K.}~\bibnamefont {Watanabe}}, \ and\ \bibinfo
  {author} {\bibfnamefont {J.}~\bibnamefont {Yan}},\ }\href {\doibase
  10.1038/s41467-018-05558-x} {\bibfield  {journal} {\bibinfo  {journal}
  {Nature Communications}\ }\textbf {\bibinfo {volume} {9}},\ \bibinfo {pages}
  {3717} (\bibinfo {year} {2018}{\natexlab{b}})}\BibitemShut {NoStop}%
\bibitem [{\citenamefont {Li}\ \emph {et~al.}(2018)\citenamefont {Li},
  \citenamefont {Wang}, \citenamefont {Lu}, \citenamefont {Jin}, \citenamefont
  {Chen}, \citenamefont {Meng}, \citenamefont {Lian}, \citenamefont
  {Taniguchi}, \citenamefont {Watanabe}, \citenamefont {Zhang}, \citenamefont
  {Smirnov},\ and\ \citenamefont {Shi}}]{Li2018}%
  \BibitemOpen
  \bibfield  {author} {\bibinfo {author} {\bibfnamefont {Z.}~\bibnamefont
  {Li}}, \bibinfo {author} {\bibfnamefont {T.}~\bibnamefont {Wang}}, \bibinfo
  {author} {\bibfnamefont {Z.}~\bibnamefont {Lu}}, \bibinfo {author}
  {\bibfnamefont {C.}~\bibnamefont {Jin}}, \bibinfo {author} {\bibfnamefont
  {Y.}~\bibnamefont {Chen}}, \bibinfo {author} {\bibfnamefont {Y.}~\bibnamefont
  {Meng}}, \bibinfo {author} {\bibfnamefont {Z.}~\bibnamefont {Lian}}, \bibinfo
  {author} {\bibfnamefont {T.}~\bibnamefont {Taniguchi}}, \bibinfo {author}
  {\bibfnamefont {K.}~\bibnamefont {Watanabe}}, \bibinfo {author}
  {\bibfnamefont {S.}~\bibnamefont {Zhang}}, \bibinfo {author} {\bibfnamefont
  {D.}~\bibnamefont {Smirnov}}, \ and\ \bibinfo {author} {\bibfnamefont
  {S.-F.}\ \bibnamefont {Shi}},\ }\href {\doibase 10.1038/s41467-018-05863-5}
  {\bibfield  {journal} {\bibinfo  {journal} {Nature Communications}\ }\textbf
  {\bibinfo {volume} {9}},\ \bibinfo {pages} {3719} (\bibinfo {year}
  {2018})}\BibitemShut {NoStop}%
\bibitem [{\citenamefont {{Molas}}\ \emph {et~al.}(2019)\citenamefont
  {{Molas}}, \citenamefont {{Slobodeniuk}}, \citenamefont {{Kazimierczuk}},
  \citenamefont {{Nogajewski}}, \citenamefont {{Bartos}}, \citenamefont
  {{Kapu{\'s}ci{\'n}ski}}, \citenamefont {{Oreszczuk}}, \citenamefont
  {{Watanabe}}, \citenamefont {{Taniguchi}}, \citenamefont {{Faugeras}},
  \citenamefont {{Kossacki}}, \citenamefont {{Basko}},\ and\ \citenamefont
  {{Potemski}}}]{molasWSe2}%
  \BibitemOpen
  \bibfield  {author} {\bibinfo {author} {\bibfnamefont {M.~R.}\ \bibnamefont
  {{Molas}}}, \bibinfo {author} {\bibfnamefont {A.~O.}\ \bibnamefont
  {{Slobodeniuk}}}, \bibinfo {author} {\bibfnamefont {T.}~\bibnamefont
  {{Kazimierczuk}}}, \bibinfo {author} {\bibfnamefont {K.}~\bibnamefont
  {{Nogajewski}}}, \bibinfo {author} {\bibfnamefont {M.}~\bibnamefont
  {{Bartos}}}, \bibinfo {author} {\bibfnamefont {P.}~\bibnamefont
  {{Kapu{\'s}ci{\'n}ski}}}, \bibinfo {author} {\bibfnamefont {K.}~\bibnamefont
  {{Oreszczuk}}}, \bibinfo {author} {\bibfnamefont {K.}~\bibnamefont
  {{Watanabe}}}, \bibinfo {author} {\bibfnamefont {T.}~\bibnamefont
  {{Taniguchi}}}, \bibinfo {author} {\bibfnamefont {C.}~\bibnamefont
  {{Faugeras}}}, \bibinfo {author} {\bibfnamefont {P.}~\bibnamefont
  {{Kossacki}}}, \bibinfo {author} {\bibfnamefont {D.~M.}\ \bibnamefont
  {{Basko}}}, \ and\ \bibinfo {author} {\bibfnamefont {M.}~\bibnamefont
  {{Potemski}}},\ }\href@noop {} {\bibfield  {journal} {\bibinfo  {journal}
  {arXiv e-prints}\ } (\bibinfo {year} {2019})},\ \Eprint
  {http://arxiv.org/abs/1901.04431} {arXiv:1901.04431 [cond-mat.mes-hall]}
  \BibitemShut {NoStop}%
\bibitem [{\citenamefont {{Liu}}\ \emph {et~al.}(2019)\citenamefont {{Liu}},
  \citenamefont {{van Baren}}, \citenamefont {{Lu}}, \citenamefont
  {{Altaiary}}, \citenamefont {{Taniguchi}}, \citenamefont {{Watanabe}},
  \citenamefont {{Smirnov}},\ and\ \citenamefont {{Lui}}}]{liuTRIONS}%
  \BibitemOpen
  \bibfield  {author} {\bibinfo {author} {\bibfnamefont {E.}~\bibnamefont
  {{Liu}}}, \bibinfo {author} {\bibfnamefont {J.}~\bibnamefont {{van Baren}}},
  \bibinfo {author} {\bibfnamefont {Z.}~\bibnamefont {{Lu}}}, \bibinfo {author}
  {\bibfnamefont {M.~M.}\ \bibnamefont {{Altaiary}}}, \bibinfo {author}
  {\bibfnamefont {T.}~\bibnamefont {{Taniguchi}}}, \bibinfo {author}
  {\bibfnamefont {K.}~\bibnamefont {{Watanabe}}}, \bibinfo {author}
  {\bibfnamefont {D.}~\bibnamefont {{Smirnov}}}, \ and\ \bibinfo {author}
  {\bibfnamefont {C.~H.}\ \bibnamefont {{Lui}}},\ }\href@noop {} {\bibfield
  {journal} {\bibinfo  {journal} {arXiv e-prints}\ } (\bibinfo {year}
  {2019})},\ \Eprint {http://arxiv.org/abs/1901.11043} {arXiv:1901.11043
  [cond-mat.mtrl-sci]} \BibitemShut {NoStop}%
\bibitem [{\citenamefont {{Goryca}}\ \emph {et~al.}(2019)\citenamefont
  {{Goryca}}, \citenamefont {{Li}}, \citenamefont {{Stier}}, \citenamefont
  {{Crooker}}, \citenamefont {{Taniguchi}}, \citenamefont {{Watanabe}},
  \citenamefont {{Courtade}}, \citenamefont {{Shree}}, \citenamefont
  {{Robert}}, \citenamefont {{Urbaszek}},\ and\ \citenamefont
  {{Marie}}}]{goryca2019}%
  \BibitemOpen
  \bibfield  {author} {\bibinfo {author} {\bibfnamefont {M.}~\bibnamefont
  {{Goryca}}}, \bibinfo {author} {\bibfnamefont {J.}~\bibnamefont {{Li}}},
  \bibinfo {author} {\bibfnamefont {A.~V.}\ \bibnamefont {{Stier}}}, \bibinfo
  {author} {\bibfnamefont {S.~A.}\ \bibnamefont {{Crooker}}}, \bibinfo {author}
  {\bibfnamefont {T.}~\bibnamefont {{Taniguchi}}}, \bibinfo {author}
  {\bibfnamefont {K.}~\bibnamefont {{Watanabe}}}, \bibinfo {author}
  {\bibfnamefont {E.}~\bibnamefont {{Courtade}}}, \bibinfo {author}
  {\bibfnamefont {S.}~\bibnamefont {{Shree}}}, \bibinfo {author} {\bibfnamefont
  {C.}~\bibnamefont {{Robert}}}, \bibinfo {author} {\bibfnamefont
  {B.}~\bibnamefont {{Urbaszek}}}, \ and\ \bibinfo {author} {\bibfnamefont
  {X.}~\bibnamefont {{Marie}}},\ }\href@noop {} {\bibfield  {journal} {\bibinfo
   {journal} {arXiv e-prints}\ ,\ \bibinfo {pages} {arXiv:1904.03238}}
  (\bibinfo {year} {2019})},\ \Eprint {http://arxiv.org/abs/1904.03238}
  {arXiv:1904.03238 [cond-mat.mes-hall]} \BibitemShut {NoStop}%
\end{thebibliography}%


\begin{thebibliography}{99}
	
	\bibitem{gomezSM}
	A. Castellanos-Gomez, M. Buscema, R. Molenaar, V. Singh, L. Janssen, H. S. J. van der Zant, and G. A. Steele, 2D Materials {\bf 1}, 011002 (2014).
	
	\bibitem{kratzerSM}
	A. Kratzer, Zeitschrift f\"ur Physik {\bf 3}, 289 (1920).
	
	\bibitem{YangSM}
	X. L. Yang, S. H. Guo, F. T. Chan, K. W. Wong, and W. Y. Ching, Phys. Rev. A {\bf 43}, 1186 (1991).
	
	\bibitem{rytovaSM}
	N. S. Rytova, Proc. MSU, Phys. Astron. {\bf 3}, 308 (1967).
	
	\bibitem{keldyshSM}
	L. V. Keldysh, JETP Lett. {\bf 29}, 658 (1979).
	
	\bibitem{betheSM}
	H. A. Bethe and E. M. Salpeter, Quantum mechanics of one and two-electron atoms (Springer-Verlag Berlin, 1957).
	
	\bibitem{stierSM}
	A. V. Stier, N. P. Wilson, K. A. Velizhanin, J. Kono, X. Xu, and S. A. Crooker, Phys. Rev. Lett. {\bf 120}, 057405 (2018).
	
	\bibitem{berkelbachSM}
	T. C. Berkelbach, M. S. Hybertsen, and D. R. Reichman, Phys. Rev. B {\bf 88}, 045318 (2013).
	
	\bibitem{batemanSM}
	H. Bateman and A. Erdelyi, Higher Transcendental Functions, Volume 1 (Mc Graw-Hill book Co. New York, 1953).
	
	\bibitem{geickSM}
	R. Geick, C. H. Perry, and G. Rupprecht, Phys. Rev. {\bf 146}, 543 (1966).
	
	\bibitem{cudazzoSM}
	P. Cudazzo, I. V. Tokatly, and A. Rubio, Phys. Rev. B {\bf 84}, 085406 (2011).
	
	\bibitem{liSM}
	Y. Li, A. Chernikov, X. Zhang, A. Rigosi, H. M. Hill, A. M. van der Zande, D. A. Chenet, E.-M. Shih, J. Hone, and T. F. Heinz, Phys. Rev. B {\bf 90}, 205422 (2014).
	
	\bibitem{aroraWSe2SM}
	A. Arora, M. Koperski, K. Nogajewski, J. Marcus, C. Faugeras, and M. Potemski, Nanoscale {\bf 7}, 10421 (2015).
	
	\bibitem{aroraMoSe2SM}
	A. Arora, K. Nogajewski, M. Molas, M. Koperski, and M. Potemski, Nanoscale {\bf 7}, 20769 (2015).
	
	\bibitem{molasSM}
	M. R. Molas, K. Nogajewski, A. O. Slobodeniuk, J. Binder, M. Bartos, and M. Potemski, Nanoscale {\bf 9}, 13128 (2017).
	
	\bibitem{robertSM}
	C. Robert, M. A. Semina, F. Cadiz, M. Manca, E. Courtade, T. Taniguchi, K. Watanabe, H. Cai, S. Tongay, B. Lassagne, P. Renucci, T. Amand, X. Marie, M. M. Glazov, and B. Urbaszek, Phys. Rev. Materials {\bf 2}, 011001 (2018).
	
	\bibitem{hanSM}
	B. Han, C. Robert, E. Courtade, M. Manca, S. Shree, T. Amand, P. Renucci, T. Taniguchi, K. Watanabe, X. Marie, L. E. Golub, M. M. Glazov, and B. Urbaszek, Phys. Rev. X {\bf 8}, 031073 (2018).
	
	\bibitem{slobodeniukSM}
	A. O. Slobodeniuk,  L. Bala, M. Koperski, M. R. Molas, P. Kossacki, K. Nogajewski, M. Bartos, K. Watanabe, T. Taniguchi, C. Faugeras, and M. Potemski, arXiv e-prints (2018), arXiv:1810.00623 [cond-mat.mes-hall].
	
	\bibitem{gerbarSM}
	I. C. Gerber, E. Courtade, S. Shree, C. Robert, T. Taniguchi, K. Watanabe, A. Balocchi, P. Renucci, D. Lagarde, X. Marie, and B. Urbaszek, Phys. Rev. B {\bf 99}, 035443 (2019).

	\bibitem{gorycaSM}	
	M. Goryca, J. Li, A. V. Stier, S. A. Crooker, T. Taniguchi, K. Watanabe, E. Courtade, S. Shree, C. Robert, B. Urbaszek, and X. Marie, arXiv e-prints , arXiv:1904.03238 (2019), arXiv:1904.03238.

	\bibitem{chernikovSM}	
	A. Chernikov, T. C. Berkelbach, H. M. Hill, A. Rigosi, Y. Li, O. B. Aslan, D. R. Reichman, M. S. Hybertsen, and T. F. 	Heinz, Phys. Rev. Lett. {\bf 113}, 076802 (2014).
	
\end{thebibliography}

\newpage
\onecolumngrid
\setcounter{figure}{0}
\setcounter{section}{0}
\renewcommand{\thefigure}{S\arabic{figure}}
\renewcommand{\thesection}{S\arabic{section}}

\begin{center}
	{\large{ {\bf Supplemental Material: \\ Energy spectrum of two-dimensional excitons in a non-uniform dielectric medium}}}
	\vskip0.5\baselineskip{M. R. Molas,{$^{1,2}$} A. O. Slobodeniuk,{$^{1}$} K. Nogajewski,{$^{1,2}$} M. Bartos,{$^{1,3}$} \L{}. Bala,{$^{1,2}$} \linebreak[4] A. Babi\'nski,{$^{2}$} K. Watanabe,{$^{4}$} T. Taniguchi,{$^{4}$} C. Faugeras,{$^{1}$} and M. Potemski,{$^{1,2}$}}
	\vskip0.5\baselineskip{\em$^{1}$ Laboratoire National des Champs Magn\'etiques Intenses, CNRS-UGA-UPS-INSA-EMFL, 25, avenue des Martyrs, 38042 Grenoble, France \\$^{2}$ Faculty of Physics, University of Warsaw, ul. Pasteura 5, 02-093 Warszawa, Poland \\$^{3}$Central European Institute of Technology, Brno University of Technology,  Purky\v{n}ova 656/123, 612 00 BRNO, Czech Republic \\$^{4}$National Institute for Materials Science, 1-1 Namiki, Tsukuba 305-0044, Japan }
\end{center}

This supplemental material provides: \ref{exp} description of preparation of the studied samples and used experimental setups, \ref{kratzer} excitonic spectrum and eigenfunctions in the Kratzer potential, \ref{keldysh} numerical analysis of the excitonic spectrum in the Rytova-Keldysh potential, \ref{spectrum} derivation of the excitonic spectrum in WSe$_2$ monolayer encapsulated in hBN, \ref{inne} magneto-photoluminescence investigation of MoS$_2$ and WS$_2$ monolayers, \ref{sigma} dependence of excitonic diamagnetic coefficients in WSe$_2$ monolayer, \ref{rc} low temperature reflectance contrast spectra of the investigated monolayers, \ref{bandgap} estimation of the band-gap energy in MoSe$2$ monolayer, \ref{test} application of the proposed model to the data available in the literature.

\section{S\lowercase{amples and experimental setups}\label{exp}}

The active parts of our samples consist of a monolayer (ML) of semiconducting transition metal
dichalcogenides (S-TMD), $i.e.$ WSe$_2$, MoS$_2$, WS$_2$, and MoSe$_2$, which has been encapsulated
in hexagonal boron nitride (hBN) and deposited on a bare Si substrate. They were fabricated by
two-stage polydimethylsiloxane (PDMS)-based~\cite{gomezSM} mechanical exfoliation of S-TMD and hBN
bulk crystals.

The encapsulating hBN layers were rather thick, the bottom hBN layers were 35-40 nm-thick in case of
WS2 and WSe2 monolayers and 60-65 nm-thick in case of MoS2 and MoSe2 monolayers. The thickness of the
top hBN was of about 20nm in all structures studied.

The $\mu$-photoluminescence ($\mu$-PL) and $\mu$-reflectance contrast ($\mu$-RC) experiments were
performed using a $\lambda$=515~nm CW laser diode and a 100~W tungsten halogen lamp, respectively.

Micro-magneto-PL measurements were performed in the Faraday configuration using an optical-fiber-based
insert placed in a superconducting magnetic coil producing magnetic fields up to 14 T. The sample
was mounted on top of an $x-y-z$ piezo-stage kept in gaseous helium at $T$= 4.2 K. The excitation
light was coupled to an optical fiber with a core of 5~$\mu$m diameter and focused on the sample
by an aspheric lens (spot diameter around 1~$\mu$m). The signal was collected by the same lens,
injected into a second optical fiber of 50~$\mu$m diameter, and analyzed by a \mbox{0.5 m} long
monochromator equipped with a CCD camera. A combination of a quarter wave plate and a polarizer are
used to analyse the circular polarization of signals. The measurements were performed with a fixed
circular polarization, whereas reversing the direction of magnetic field yields the information
corresponding to the other polarization component due to time-reversal symmetry.

Investigations at zero magnetic field were carried out with the aid of a continuous flow cryostat
mounted on $x-y$ motorized positioners. The sample was placed on a cold finger of the cryostat.
The excitation light was focused by means of a 50x long-working distance objective with a 0.5 numerical
aperture producing a spot of about 1~$\mu$m. The signal was collected via the same microscope objective,
sent through a 0.5 m monochromator, and then detected by a CCD camera.

\section{E\lowercase{xcitonic spectrum and eigenfunctions in the} K\lowercase{ratzer potential}\label{kratzer}}

We solve two-dimensional (2D) Schr\"{o}dinger equation with the Kratzer potential~\cite{kratzerSM}
for wave-function $\psi(\mathbf{r})=\psi(r,\varphi)$
\begin{equation}
\Big\{-\frac{\hbar^2}{2\mu}\Big[\frac{\partial^2}{\partial r^2}+\frac{1}{r}\frac{\partial}{\partial r}+
\frac{1}{r^2}\frac{\partial^2}{\partial\varphi^2}\Big]+U_{ext}(r) - \epsilon\Big\}\psi(r,\varphi)=0,
\end{equation}
in which $U_{ext}(r)=-e^2/r_0(r_0^*/r-g^2r_0^{*2}/r^2)$ is the modified Kratzer potential. $r$ is in-plane electron-hole distance, $\mu$~denotes the reduced
electron-hole mass, $\varepsilon$ represents the dielectric constant of the material surrounding
the monolayer, $r_0^*=r_0/\varepsilon$ is the reduced screening length, and $g$ is a tunable parameter.
Taking $\psi_m(r,\varphi)=e^{im\varphi}\phi_m(r)/\sqrt{2\pi}$ and introducing the new
variable $\xi=r/r_0^*$, we obtain the equation
\begin{equation}
\Big\{\frac{d^2}{d\xi^2}+\frac{1}{\xi}\frac{d}{d\xi}+
\frac{-k^2\xi^2+\kappa^2\xi-g^2\kappa^2-m^2}{\xi^2}\Big\}\phi_m(\xi)=0
\end{equation}
with $k^2=-2\mu\epsilon r_0^{*2}/\hbar^2>0$ and $\kappa^2=2\mu r_0e^2/\hbar^2\varepsilon^2>0$.
The solution to this eigenvalue problem is
\begin{equation}
\label{eq:full_wave_function}
\phi_{n,m}(r)=\frac{\beta_{n,m}}{\sqrt{2n+2\delta_m-1}}\sqrt{\frac{(n-|m|-1)!}{\Gamma(n+|m|+2\delta_m)}}\times (\beta_{n,m}r)^M e^{-\beta_{n,m}r/2}
L_{n-|m|-1}^{2M}(\beta_{n,m}r),
\end{equation}
with $M=\sqrt{m^2+g^2\kappa^2}$, $\delta_m=M-|m|$ and $\beta_{n,m}=2\mu e^2/\hbar^2\varepsilon(n+\delta_m-1/2)$,
respectively. $n$=1, 2\dots~is a principal quantum number, $m$=0, $\pm1$, $\pm2$\dots~is an angular momentum quantum number, and
$L_n^\alpha(x)$ is the modified Laguerre polynomial. The energy spectrum for such system is described with:
\begin{equation}
\label{eq:full_spectrum}
\epsilon_{n,m}=-\frac{\mu e^4}{2\hbar^2\varepsilon^2}\frac{1}{(n+\delta_m-\frac12)^2}.
\end{equation}
For $g$=0, our result coincides with 2D hydrogen model~\cite{YangSM}. In the case of $s$-type states
($n$=1, 2\dots~and $m$=0), the excitonic spectrum simplifies to
\begin{equation}
\label{eq:partial_spectrum}
\epsilon_n=-\frac{\mu e^4}{2\hbar^2\varepsilon^2}\frac{1}{(n+g\kappa-\frac12)^2},
\end{equation}
We mention the following consequences of this model:
(i) the energy scale (prefactor in $\epsilon_n$) does not depend on the screening length $r_0$. It coincides with the Rydberg constant $Ry^*$ for an exciton with reduced mass $\mu$ in an environment
with dielectric constant $\varepsilon$, $i.e.$ $Ry^*=\mu e^4/2\hbar^2\varepsilon^2$;
(ii) the information about relative positions of the energy levels of the system is encoded in the denominators in Eq.~\ref{eq:full_spectrum} and \ref{eq:partial_spectrum};
(iii) the Kratzer potential lifts the Coulomb degeneracy of the $s$- ($m$=0) and $p$-type ($m$=$\pm$1) states, as one can be noticed from Eq.~(\ref{eq:full_spectrum});
(iv)  since  $Ry^*\propto 1/\varepsilon^2$ and $\delta+1/2\propto 1/\varepsilon$, the energy ladder of the excitons can be progressively tuned by changing the dielectric constant $\varepsilon$ of the surrounding medium. Surprisingly, the results of numerical simulations performed in the Rytova-Keldysh potential~\cite{rytovaSM,keldyshSM}, discussed in the next section, demonstrate the similar behavior. This fact can be interpreted as the indirect confirmation, that the Kratzer potential is a good approximation for the considered model.

Using the wave-functions obtained in Eq. (\ref{eq:full_wave_function}), we calculate the mean value
of $r^2$, which can be useful for analysis of diamagnetic shift of excitons, which reads
\begin{equation}
\langle r^2\rangle_{n,m}=\frac{2}{(\beta_{n,m})^2}[3-3m^2+5n(n-1)-5\delta_m-6|m|\delta_m+2\delta_m(5n+\delta_m)].
\end{equation}
For $s$-type states characterized by $m$=0, it takes the form
\begin{equation}
\langle r^2\rangle_{n,0}=\frac{2}{(\beta_{n,0})^2}(2g^2\kappa^2+10g\kappa n-5g\kappa+5n^2-5n+3).
\end{equation}
Moreover, for the special case $g\kappa=1/2$, while Eq.~\ref{eq:partial_spectrum} resembles the three-dimensional (3D) hydrogen model,
the mean value of the $r^2$ parameter is given by
\begin{equation}
\langle r^2\rangle_{n,0}|_{g\kappa=1/2}=(a_0^*)^2n^2(5n^2+1)/2\approx 5(a_0^*)^2n^4/2,
\label{eq:diamag}
\end{equation}
where $a_0^*$=$\hbar^2\varepsilon/\mu e^2$  is the effective Bohr radius.
It is interesting to note that the latter formula coincides with the mean value of $r^2$ for 3D hydrogen atom~\cite{betheSM}.

The eigenfunctions of $s$-states of the Schr\"{o}dinger Hamiltonian with the Kratzer potential (\ref{eq:full_wave_function}) tend to
zero at $r\rightarrow 0$. This is the consequence of the repulsive part of the potential at short distances. Therefore,
such solutions can not be good approximation for $s$-state exciton wave-functions at small distances, since the Rytova-Keldysh
potential is attractive. In order to improve the current result, one needs to modify the Kratzer potential at small distances.

\section{N\lowercase{umerical analysis of the excitonic spectrum in the} R\lowercase{ytova-}K\lowercase{eldysh potential}\label{keldysh}}
\label{sec:S3}
\begin{figure}[t]
	\includegraphics[width=8cm]{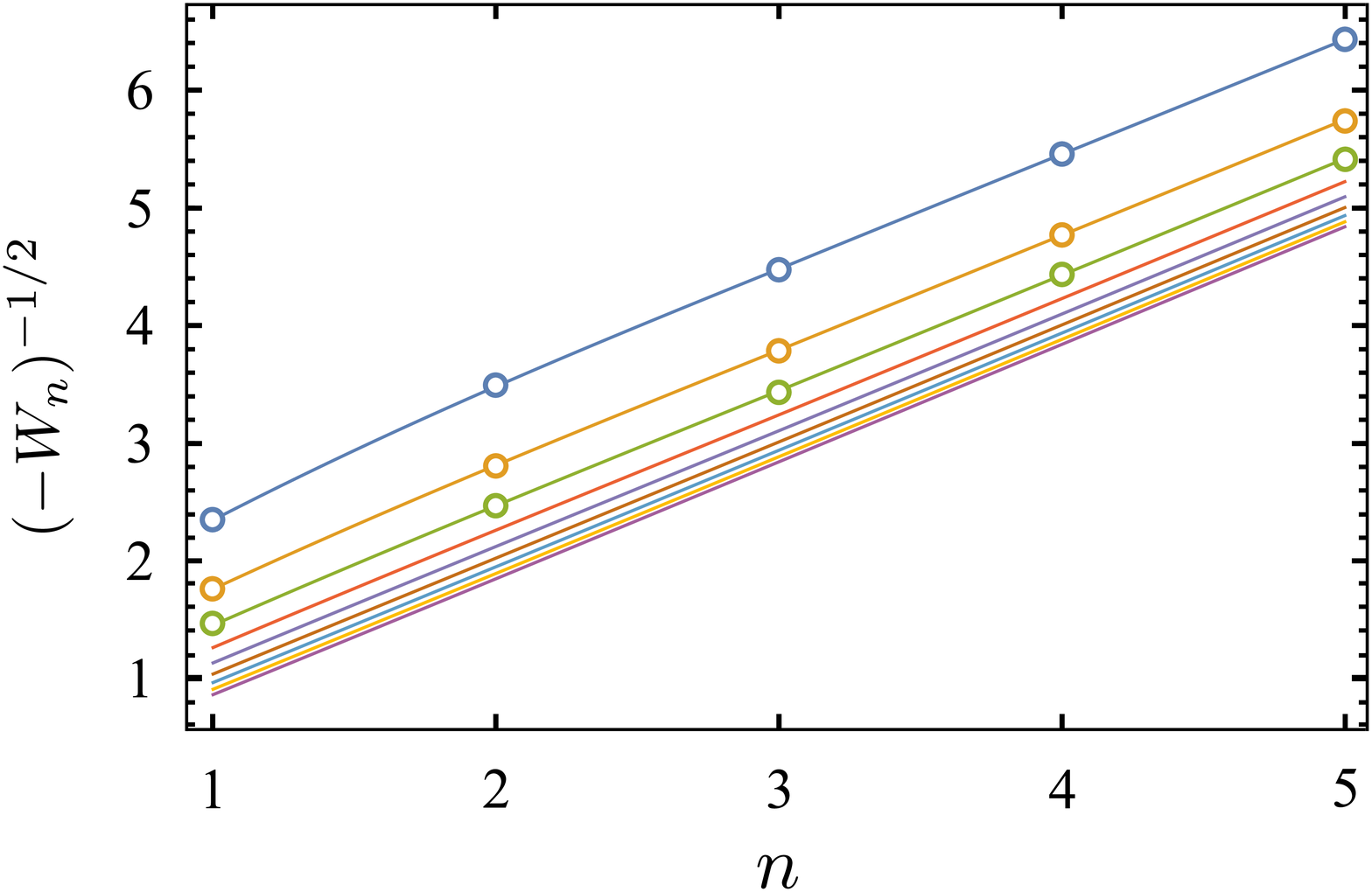}
	\caption{\label{fig:spectrum_RK}
		Interpolation lines for $(-W_n)^{-1/2}$ for $n$=1, 2,\dots, 5 and different values of the dielectric constant from $\varepsilon_{min}=1$ (top blue line) until
		$\varepsilon_{max}=5$ (bottom purple line) with step $\Delta\varepsilon=0.5$. Blue, yellow, and green circles represent the values of function $(-W_n)^{-1/2}$ for different $n$,
		when $\varepsilon$=1, 1.5, 2, respectively. }
\end{figure}

We solve numerically the eigenvalue problem for 2D Schr\"{o}dinger Hamiltonian with the Rytova-Keldysh potential~\cite{rytovaSM,keldyshSM}.
We analyse the scale laws for the spectrum both as a function of principal number $n$ and dielectric constant of surrounding
medium $\varepsilon$. We start from the radial equation on wave function $\phi(r)$ for the $s$-type states characterized
by zero angular momentum:
\begin{equation}
\Big\{-\frac{\hbar^2}{2\mu}\frac{1}{r}\frac{d}{dr}\Big[r\frac{d}{dr}\Big]-\frac{\pi e^2}{2 r_0}\Big[\text{H}_0\Big(\frac{r\varepsilon}{r_0}\Big)-Y_0\Big(\frac{r\varepsilon}{r_0}\Big)\Big] - \epsilon\Big\}\phi(r)=0,
\end{equation}
where $\mathrm{H}_0(x)$ and $Y_0(x)$ are the zeroth order Struve and Neumann functions. Introducing new variables
$\xi=r\varepsilon/r_0=r/r_0^*$ and $\epsilon=(\mu e^4/2\hbar^2\varepsilon^2)W=WRy^*$, we rewrite the equation
\begin{equation}
\Big\{-b^2\frac{1}{\xi}\frac{d}{d\xi}\Big[\xi\frac{d}{d\xi}\Big]-\pi b\Big[\text{H}_0(\xi)-Y_0(\xi)\Big] - W\Big\}\phi(\xi)=0
\label{shred}
\end{equation}
with $b=\hbar^2\varepsilon^2/(\mu e^2r_0)=a_B^*/r_0^*$ -- the ratio of the natural scales in the system.
We derive the spectrum of this differential equation as a function of $\varepsilon$ and $n$.

\begin{figure}[b]
	\includegraphics[width=8cm]{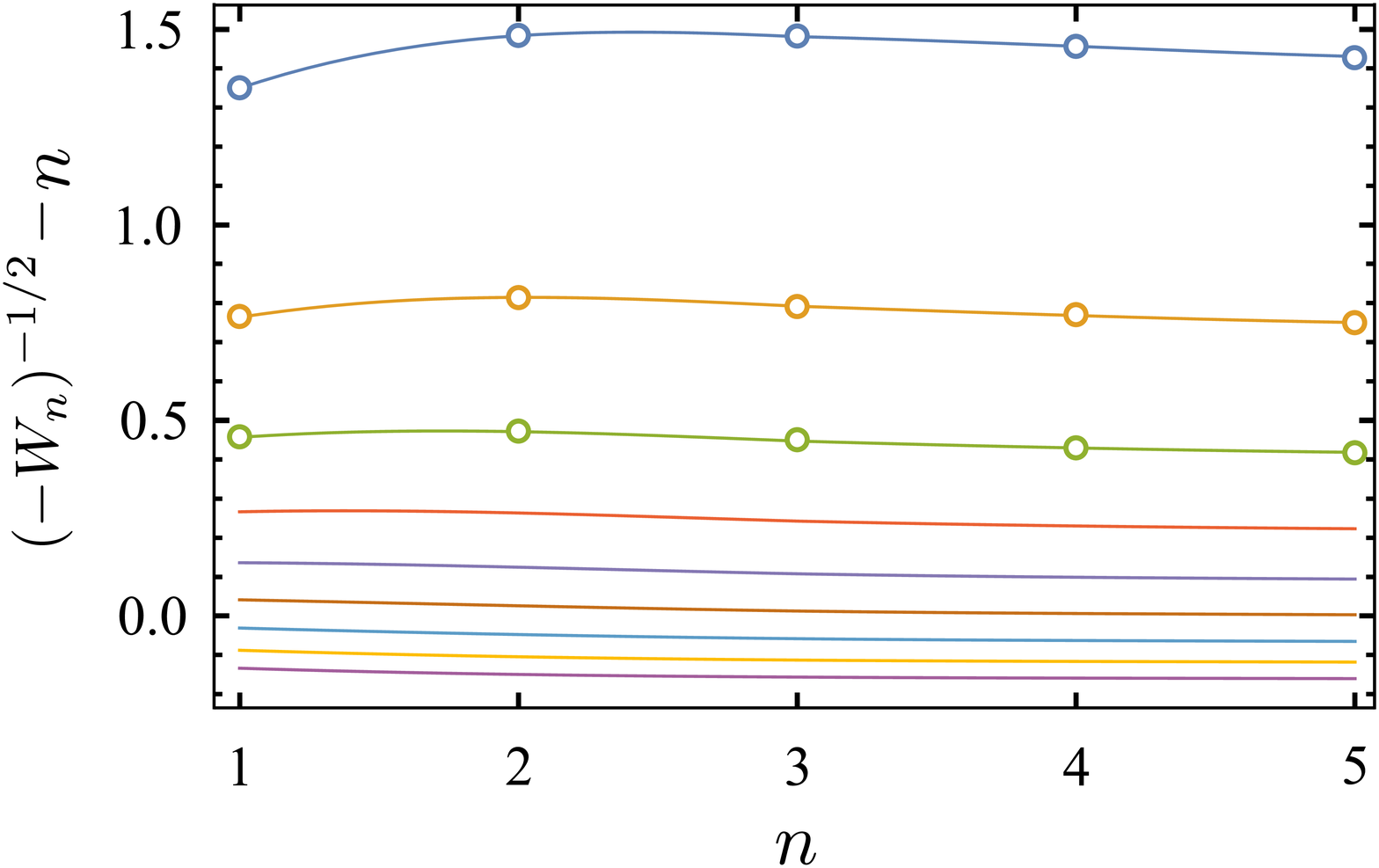}
	\caption{\label{fig:spectrum_RK_diff}
		Interpolation lines for $(-W_n)^{-1/2}-n$ for $n$=1, 2, \dots, 5 and different values of the dielectric constant from $\varepsilon_{min}=1$ (top blue line) to
		$\varepsilon_{max}=5$ (bottom purple line) with step $\Delta\varepsilon=0.5$. Blue, yellow, and green circles represent the values of function $(-W_n)^{-1/2}-n$ for different $n$,
		when $\varepsilon$=1, 1.5, 2, respectively.}
\end{figure}

According to our hypothesis described in the main text, the excitonic spectrum should be described with
\begin{equation}
\label{eq:energy_ladder}
\epsilon_n=Ry^*W_n=-\frac{Ry^*}{(\alpha n+\beta)^2}=-Ry^*\frac{\gamma}{(n+\delta)^2},
\end{equation}
where $\gamma=\alpha^{-2}\simeq1$, while $\beta$ (and hence $\delta=\beta/\alpha$) strongly depends on $\varepsilon$.
Therefore, we estimate the linear behaviour of $(-W_n)^{-1/2}$ with $n$.

The results of numerical simulations (performed in "Mathematica") for the case of WSe$_2$ ($\mu$=0.2~$m_0$~\cite{stierSM}, $r_0=45$~$\mbox{{\AA}}$~\cite{berkelbachSM})
are presented in Figs \ref{fig:spectrum_RK} and \ref{fig:spectrum_RK_diff}. Indeed, the linear growth of $(-W_n)^{-1/2}$
as a function of $n$ for different values of $\varepsilon$ can be appreciated in Fig.~\ref{fig:spectrum_RK}.
The Fig. \ref{fig:spectrum_RK_diff} qualitatively confirms that $\alpha\simeq1$. The precision of this result
(relative deviation of the curve $(-W_n)^{-1/2}-n$ from its average value) becomes higher for larger values of $\varepsilon$.

Let us unify the aforementioned result for the case of other S-TMD monolayers and determine the limits of the applicability of our model.
	First, we derive numerically the spectrum $W_n\,(n=1,2\dots 5)$ from Eq.~\ref{shred} for different values of the parameter $b$.
	Then we fit the obtained data with the formula $\mathcal{W}_n=-\gamma/(n+\delta)^2$ and extract
	the parameters $\gamma$ and $\delta$ for each $b$.
	
\begin{figure}[t]
	\includegraphics[width=8cm]{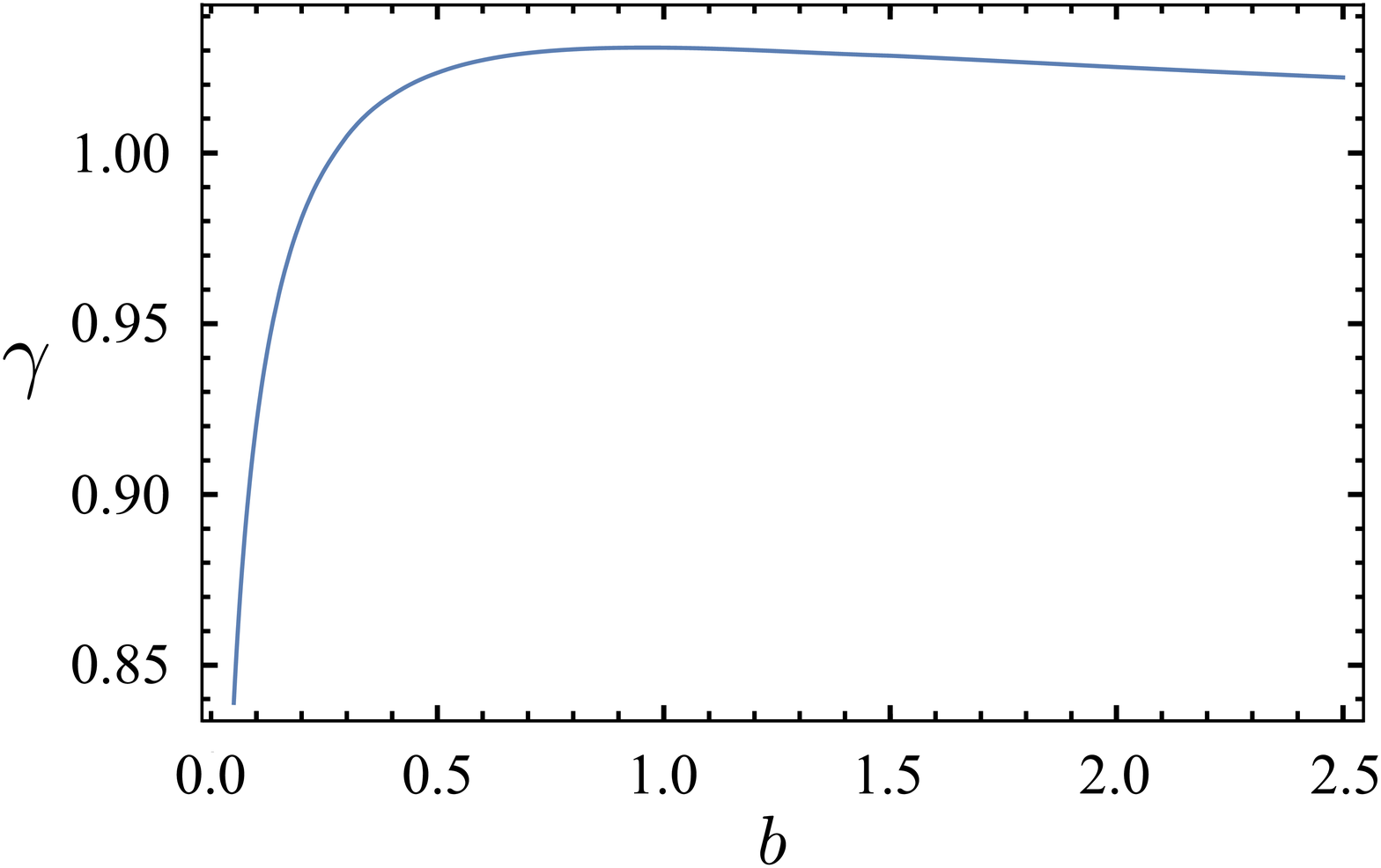}
	\caption{\label{fig:gamma}
		The parameter $\gamma$, extracted from $\mathcal{W}_n\,(n=1,2\dots5)$ for different values of the parameter $b$. }
\end{figure}

\begin{figure}[b]
	\includegraphics[width=8cm]{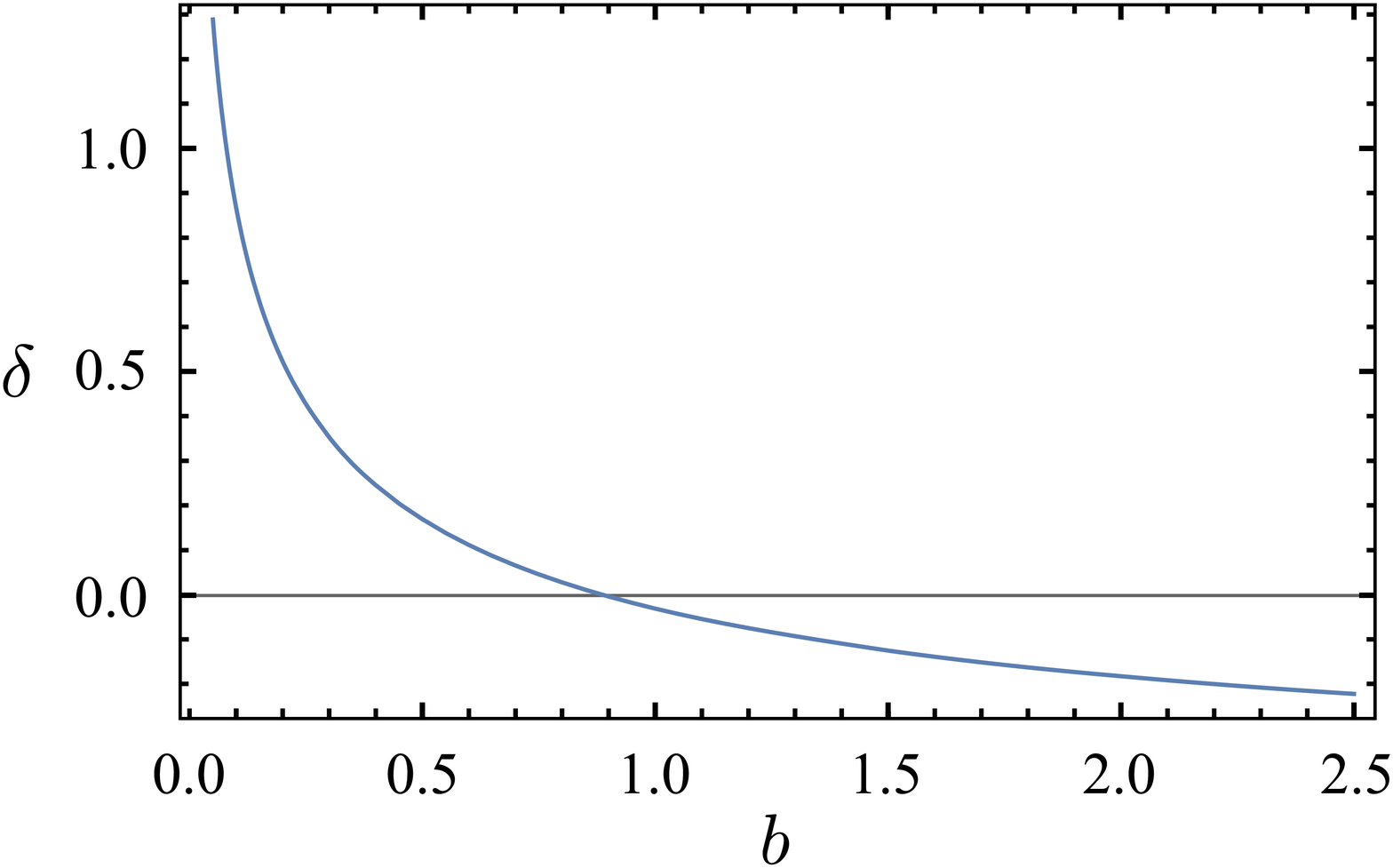}
	\caption{\label{fig:delta}
		The parameter $\delta$, extracted from $\mathcal{W}_n\,(n=1,2\dots5)$ for different values of the parameter $b$. }
\end{figure}

In further, we focus mainly on the range of parameters $b\in[0.05,2.5]$. Such a domain contains all $b$, accessible in the experiment,
	for all S-TMD monolayers and/or average dielectric constant of surrounding medium (such as PDMS, sapphire, Si/SiO$_2$ or hBN).
	The parameters $\gamma$ and $\delta$, as functions of $b$, are presented in Figs. \ref{fig:gamma} and \ref{fig:delta} respectively.

Note that the values of $\gamma$ slightly deviate from $1$ at $b>0.3$ and approach the limit $\gamma_\infty=1$ at $b\rightarrow\infty$.
	Therefore, at the relatively large $b$ the energy ladder of our problem coincides with
	the experimentally observed spectrum of excitons $\epsilon_n \approx -Ry^*/(n+\delta)^2$.
	
The values of $\delta$ are represented by monotonically decreasing function, with the limit $\delta_\infty=-1/2$ at $b\rightarrow\infty$,
	which is nothing but the case of 2D hydrogen atom. It is interesting to note that for $b\approx 0.9$ the parameter $\delta$ becomes
	zero and the exciton spectrum reproduces 3D hydrogen atom energy ladder.
	
\begin{figure}
	\includegraphics[width=8cm]{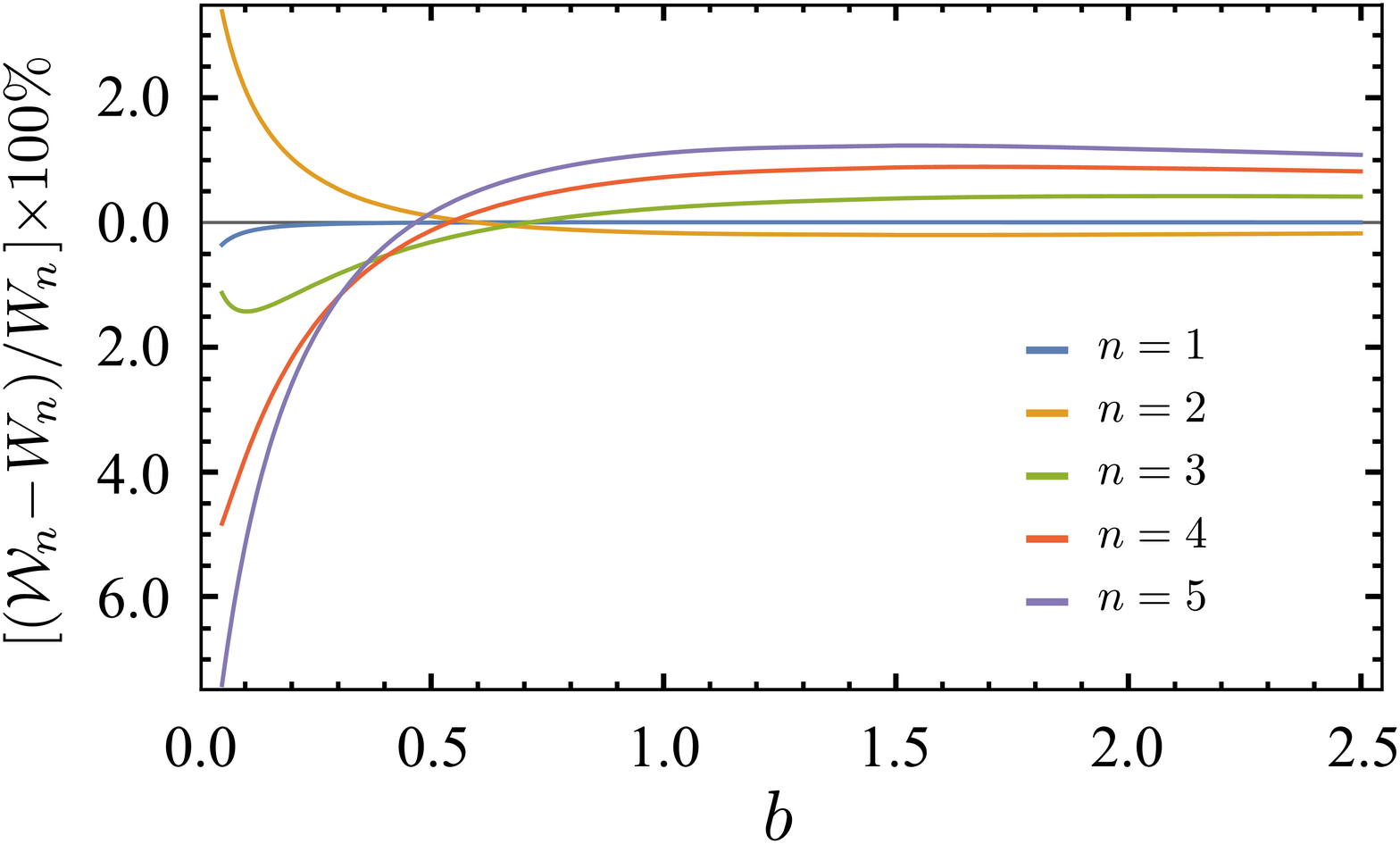}
	\caption{\label{fig:deviation}
		The relative deviation (in $\%$) between the spectrum $W_n$ and its fit $\mathcal{W}_n$ as a function of the parameter $b$ for $n=1,2\dots5$. }
\end{figure}

In order to define the limits of the applicability of our model, we calculate the relative deviations $[(\mathcal{W}_n-W_n)/W_n]\times 100\%$
	for $n=1,2,\dots 5$ as a function of parameter $b$. The corresponding plots are presented in Fig.\ref{fig:deviation}. One can see that
	starting from $b>0.3$ all the relative deviations become smaller than $2\%$ and tend to zero value at $b\rightarrow\infty$.
	The variational procedure, described in the next chapter of Supplementary Materials, predicts the ground state energy of 2D exciton in the Rytova-Keldysh potential
	with the same precision. Therefore, in order to have the self-consistent picture of all our calculations we choose the ``$2\%$ deviation rule'' as a formal criterium,
	which defines the limits of the applicability of our model.

Finally, let us discuss the specific symmetry of the studied Hamiltonian
	\begin{equation}
	H(\mu,\varepsilon,r_0)=-\frac{\hbar^2}{2\mu}\Delta_{2D}-\frac{\pi e^2}{2 r_0}\Big[\text{H}_0\Big(\frac{r\varepsilon}{r_0}\Big)-Y_0\Big(\frac{r\varepsilon}{r_0}\Big)\Big],
	\end{equation}
	where $\Delta_{2D}$ is a two-dimensional Laplacian. The Hamiltonian is a homogeneous function $H(\lambda\mu,\lambda\varepsilon,\lambda r_0)=\lambda^{-1}H(\mu,\varepsilon,r_0)$
	of its parameters $\{\mu,\varepsilon,r_0\}$ for any $\lambda>0$.
	Therefore the spectrum of the problem should have the same property: $\epsilon_n(\lambda\mu,\lambda\varepsilon,\lambda r_0)=\lambda^{-1}\epsilon_n(\mu,\varepsilon,r_0)$.
	One can demonstrate that our expression for the energy ladder (Eq.~\ref{eq:energy_ladder}) satisfies this scaling law too.

\section{D\lowercase{erivation of the excitonic spectrum in} WS\lowercase{e}$_2$ \lowercase{monolayer encapsulated in h}BN\label{spectrum}}

We consider the solution of 2D Schr\"{o}dinger equation in the potential, defined in the main text, which reads
\begin{equation}
U_{app}(\xi)=\left\{\begin{array}{cc}
-U_0\Big[\frac{1}{\xi}-\frac{0.21}{\xi^2}\Big], & \text{for}\quad \xi>\xi_0;  \\
-1.71134\,U_0, & \text{for}\quad \xi<\xi_0.
\end{array}\right.
\end{equation}
We restrict our consideration only to the $s$-type states characterized by zero angular momentum. The regular radial
solution of the Schr\"{o}dinger equation in the region $\xi<\xi_0$ is
\begin{equation}
\phi_1(\xi)\sim J_0\big(\kappa\sqrt{v_0-|\mathcal{E}|}\,\xi\big),
\end{equation}
where $J_0(x)$ is the zero-order Bessel function of the first kind, $\mathcal{E}=\epsilon/U_0$ and $v_0=1.71134$.
The solution for the region $\xi>\xi_0$ has the form
\begin{equation}
\phi_2(\xi)\sim e^{-k\xi}\xi^{g\kappa}\Psi\Big(-\frac{\kappa^2}{2k}+g\kappa+\frac12,1+2g\kappa; 2k\xi\Big)
\end{equation}
with $g^2$=0.21, $k^2=-2\mu\epsilon r_0^{*2}/\hbar^2>0$ and $\Psi(a,c;z)$ is the Tricomi's function, which is regular at $\xi\rightarrow\infty$
and solves the degenerate hypergeometric equation \cite{batemanSM}
\begin{equation}
\Big\{z\frac{d^2}{dz^2}+(c-z)\frac{d}{dz}-a\Big\}\Psi(a,c;z)=0.
\end{equation}

\begin{figure}[b]
	\includegraphics[width=8.5cm]{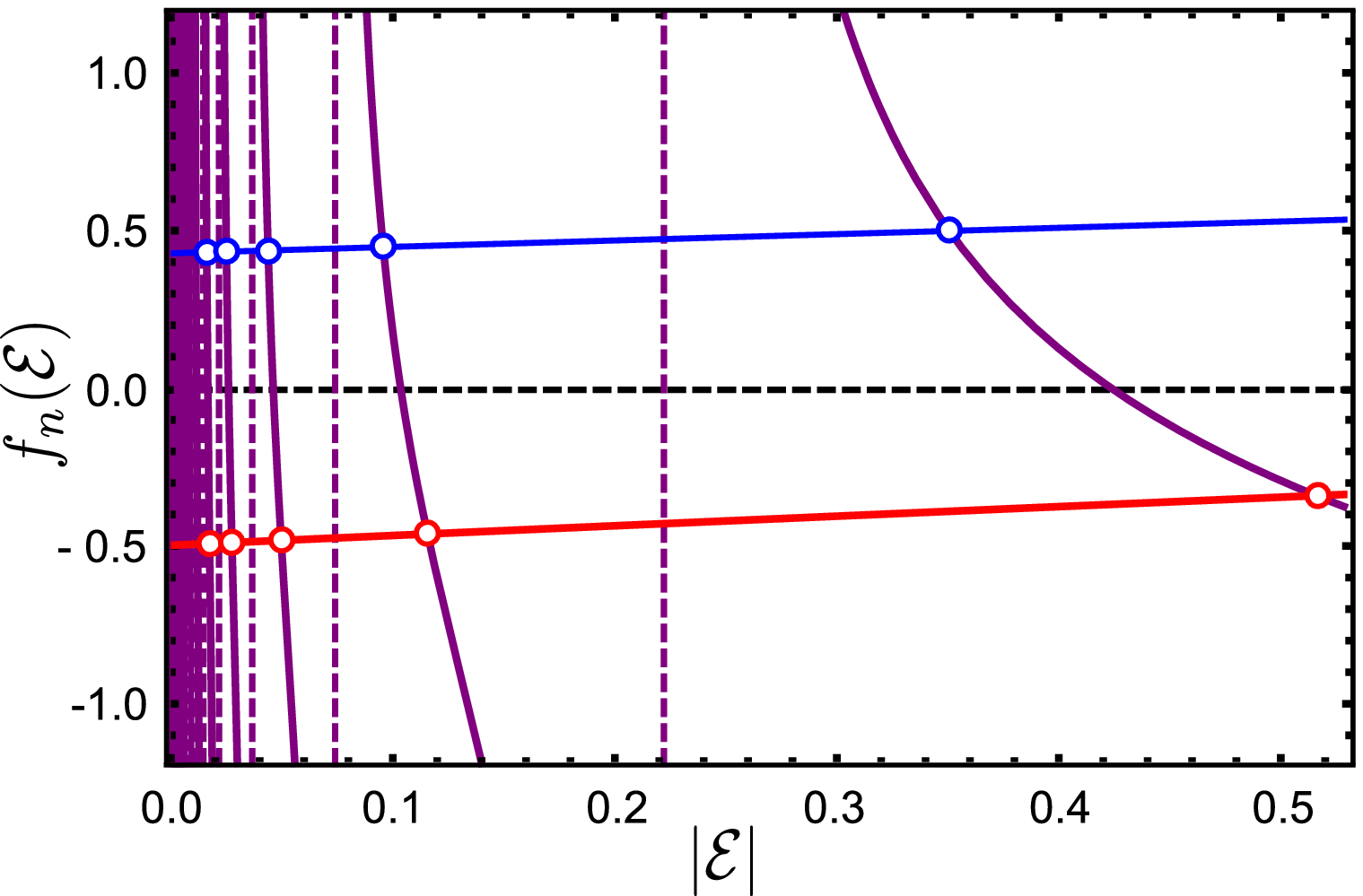}
	\caption{\label{fig:4}
		The normalized logarithmic derivatives for core solution $f_1(\mathcal{E})$ (red curve),
		external solution $f_2(\mathcal{E})$ (purple curve) and for regular solution $f_3(\mathcal{E})$ for the Kratzer potential with $g^2=0.21$ (blue curve)
		as a function of dimensionless parameter $|\mathcal{E}|=|\epsilon|/U_0$. The dashed purple lines represent the asymptotes of $f_2(\mathcal{E})$.}
\end{figure}

Introducing the normalized logarithmic derivatives for both solutions $f_n(\mathcal{E})=\kappa^{-1}[d\ln\phi_n(\xi)/d\xi]_{\xi=\xi_0},\, n=1,2$
\begin{equation}
f_1(\mathcal{E})=-\sqrt{v_0-|\mathcal{E}|}\,\,\frac{J_1(2g^2\kappa\sqrt{v_0-|\mathcal{E}|})}
{J_0(2g^2\kappa\sqrt{v_0-|\mathcal{E}|})},
\end{equation}
\begin{equation}
f_2(\mathcal{E})=-\sqrt{|\mathcal{E}|}+\frac{1}{2g}+
\big[\kappa-(2g\kappa+1)\sqrt{|\mathcal{E}|}\big]\times \frac{\Psi\Big(-\frac{\kappa}{2\sqrt{|\mathcal{E}|}}+g\kappa+\frac32,2+2g\kappa;
	4g^2\kappa\sqrt{|\mathcal{E}|}\Big)}{\Psi\Big(-\frac{\kappa}{2\sqrt{|\mathcal{E}|}}+g\kappa+\frac12,1+2g\kappa; 4 g^2\kappa\sqrt{|\mathcal{E}|}\Big)},
\end{equation}
one derives the energy spectrum from continuity equation $f_1(\mathcal{E})=f_2(\mathcal{E})$. We solve the latter equation
for WSe$_2$ ML encapsulated in hBN with a set of parameters: $\mu$=0.2~$m_0$\cite{stierSM}, $\varepsilon=4.5$~\cite{geickSM},
and $r_0=45\,\mbox{{\AA}}$~\cite{berkelbachSM}. The derivatives as a function of $|\mathcal{E}|$ are presented in Fig.~\ref{fig:4}
and the excitonic spectrum is defined by the intersection points of $f_1(\mathcal{E})$ and $f_2(\mathcal{E})$ curves.

The excitonic spectrum, obtained within the described above method, can be presented in the same form as for the Kratzer potential
(compare with Eq.~5 in the main text) and is given by
\begin{equation}
\epsilon_n=-\frac{134\,\text{meV}}{(n-0.099)^2}.
\end{equation}
This result demonstrates the good coincidence with the excitonic spectrum reported in Ref.~\citenum{stierSM}, with relative errors $8\%$ for $n=1$, $3.5\%$
for $n=2$ and less than $2\%$ for higher excited states. In order to check the precision of the graphical method, we applied it for the case, when
the core potential is described by the Kratzer one with the same parameter $g^2=0.21$. For the Kratzer potential, graphical solution provides the excitonic spectrum, which
follows Eq.~\ref{eq:partial_spectrum} of the SM.
In this case, the logarithmic derivative is
\begin{equation}
f_3(\mathcal{E})=-\sqrt{|\mathcal{E}|}+\frac{1}{2g}-
\frac{\big[\kappa-(2g\kappa+1)\sqrt{|\mathcal{E}|}\big]}{1+2g\kappa}\times \frac{_1F_1\Big(-\frac{\kappa}{2\sqrt{|\mathcal{E}|}}+g\kappa+\frac32,2+2g\kappa;
	4g^2\kappa\sqrt{|\mathcal{E}|}\Big)}{_1F_1\Big(-\frac{\kappa}{2\sqrt{|\mathcal{E}|}}+g\kappa+\frac12,1+2g\kappa; 4 g^2\kappa\sqrt{|\mathcal{E}|}\Big)},
\end{equation}
where $_1F_1(a,c;z)$ is the confluent hypergeometric function of the first kind, and corresponds to the blue curve in Fig.~\ref{fig:4}.

One can mention, that the calculation with the potential $U_{app}(\xi)$ does not approximate perfectly the $1s$-exciton state.
From the technical point of view such discrepancy can be the consequence of the modification of the Rytova-Keldysh potential
at small distances. In order to check this hypothesis, we estimate the ground state energy of exciton using another method.
Namely, we derive the ground state energy of S-TMD excitons using the Ritz variational procedure. We take the variational
wave-function in the form $\psi_0(\mathbf{r})=\beta\exp(-\beta r/2)/\sqrt{2\pi}$ and evaluate the average of the Hamiltonian
\begin{equation}
H=-\frac{\hbar^2}{2\mu}\Delta_{2D}-\frac{\pi e^2}{2 r_0}\Big[\text{H}_0\Big(\frac{r\varepsilon}{r_0}\Big)-Y_0\Big(\frac{r\varepsilon}{r_0}\Big)\Big].
\end{equation}
The kinetic energy can be calculated directly with
\begin{equation}
T_0(\beta)=-\frac{\hbar^2}{2\mu}\int_0^\infty\!\! dr\,\psi_0(\mathbf{r})\frac{d}{dr}\Big[r\frac{d\psi_0(\mathbf{r})}{dr}\Big]=\frac{\hbar^2\beta^2}{8\mu}.
\end{equation}
To determine the potential energy, a few steps need to be performed. First, we present the Rytova-Keldysh potential in the integral
form~\cite{CudazzoSM}
\begin{equation}
U(r)=-\frac{e^2}{2\pi\varepsilon}\int_0^{2\pi} d\theta\int_0^\infty dk \frac{e^{ikr\cos\theta}}{1+kr_0^*},
\end{equation}
Then, we substitute it in the expression for the average potential energy
\begin{equation}
U_0(\beta)=-\frac{e^2\beta^2}{2\pi\varepsilon}\int_0^\infty\!\!\!\frac{dk}{1+kr_0^*}\int_0^{2\pi}\!\!d\theta
\int_0^\infty\!\! drre^{(ik\cos\theta-\beta)r}.
\end{equation}
Consequently, after evaluation of integrals and adding the kinetic energy part, we determine formula
\begin{equation}
\epsilon(a)=\frac{e^2}{r_0}\Big[\frac{\hbar^2\varepsilon^2}{8\mu r_0 e^2}a^2+af(a)\Big],
\end{equation}
where $a=\beta r^*_0$ is the dimensionless parameter and
\begin{equation}
f(a)=\frac{(a-1)\sqrt{1+a^2}-2a^2\text{Arcoth}\big(\frac{1+a}{\sqrt{1+a^2}}\big)}{(1+a^2)^{3/2}}.
\end{equation}
The minimum of $\epsilon(a)$ can be found straightforwardly (with the help of "Mathematica", for example). For the case
of $\mu=0.2$~$m_0$~\cite{stierSM}, we got the ground state energy of exciton, $\epsilon_0$=-157~meV, which is in good
agreement with the results of numerical simulation reported in Ref.~\citenum{stierSM}, obtained for the same values of
parameters. Moreover, the relative deviation of the variational exciton ground-state energy, calculated for the different
values of $\varepsilon$, deviates from the numerical results discussed in Ref.~\citenum{stierSM} less than $2\%$.

Surprisingly, the numerical simulations for WSe$_2$ monolayer encapsulated in hBN with an effective mass $\mu$=0.21~$m_0$
are in better agreement with the experimentally obtained excitonic spectrum,$\epsilon_n=-140.5\,\mbox{meV}/(n-0.083)^2$
(see the main text),  than for $\mu$=0.20~$m_0$ and leads to the energy ladder of excitons given by
\begin{equation}
\epsilon_n=-141\,\mbox{meV}/(n-0.087)^2,
\end{equation}
The excitonic binding energy, $E_b$=162~meV, calculated with the aid of the Ritz variational method also nicely matches to
the experimental one, $E_b$=167~meV, obtained in the main text.

\section{M\lowercase{agneto-photoluminescence investigation of} M\lowercase{o}S$_2$ \lowercase{and} WS$_2$ \lowercase{monolayers}\label{inne}}

\begin{center}
	\begin{figure*}[h]
		\begin{minipage}[c]{0.25\linewidth}
			\includegraphics[width=1\linewidth]{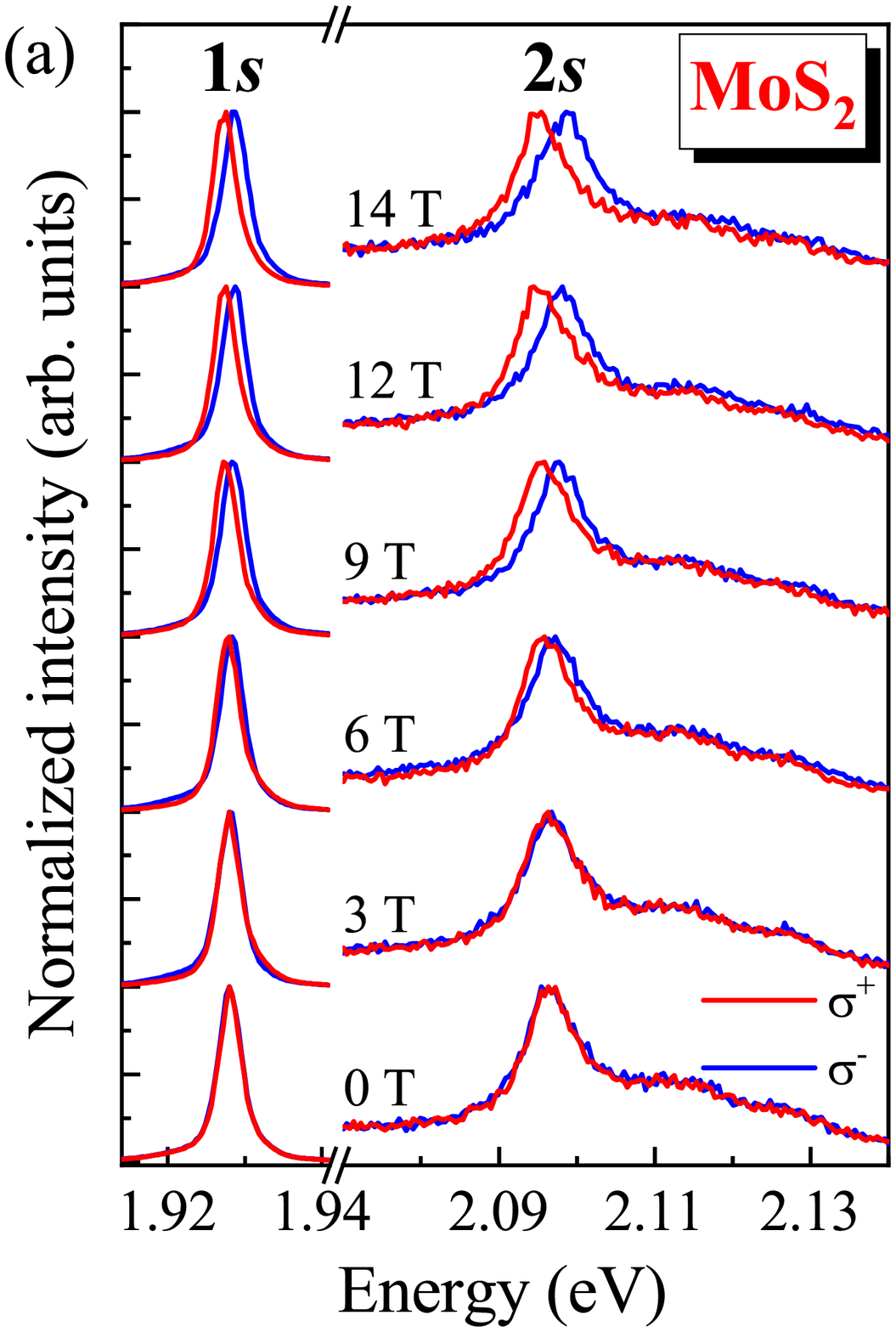}%
		\end{minipage}\hfill
		\begin{minipage}[c]{0.25\linewidth}
			\includegraphics[width=1\linewidth]{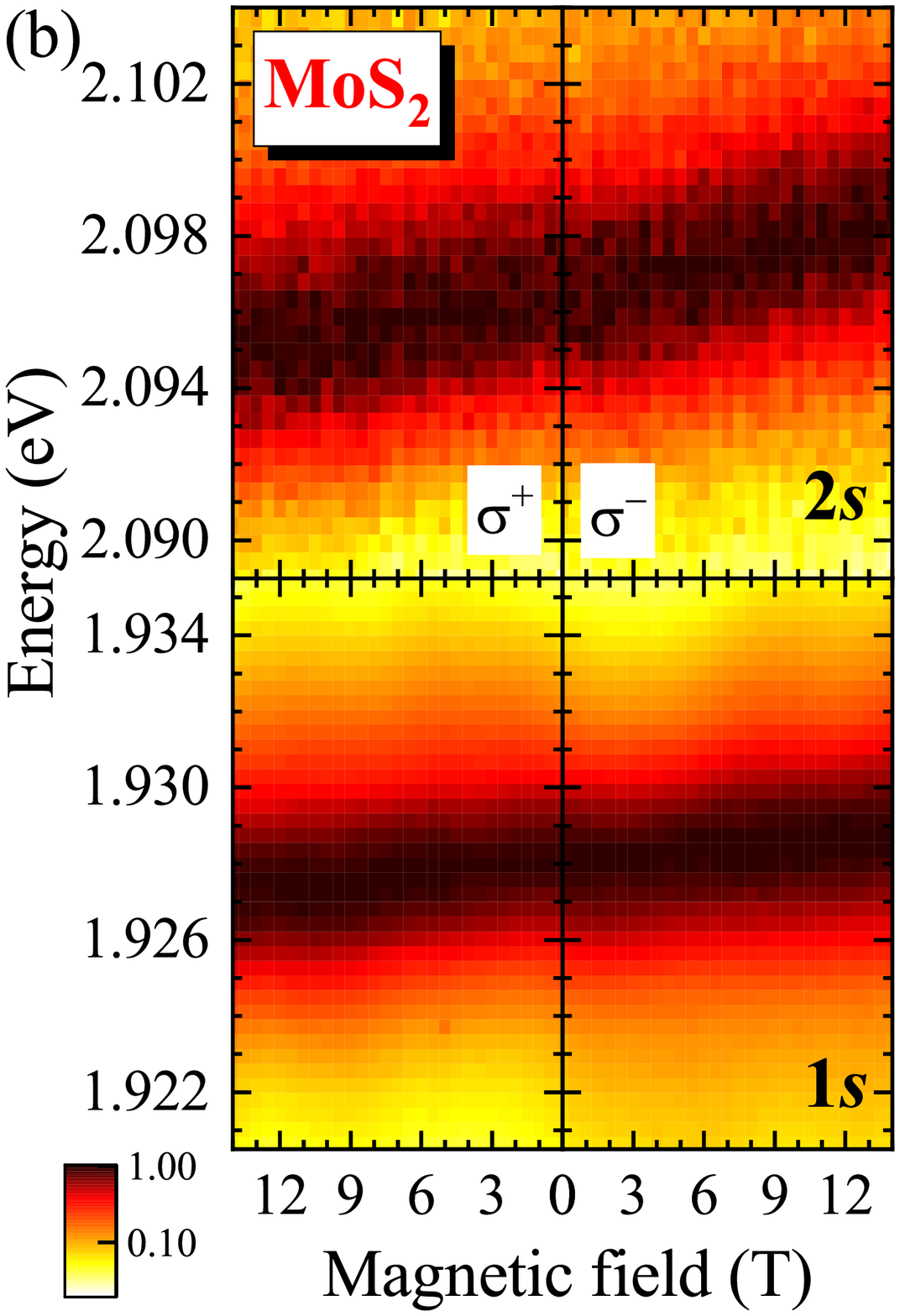}%
		\end{minipage}\hfill
		\begin{minipage}[c]{0.25\linewidth}
			\includegraphics[width=1\linewidth]{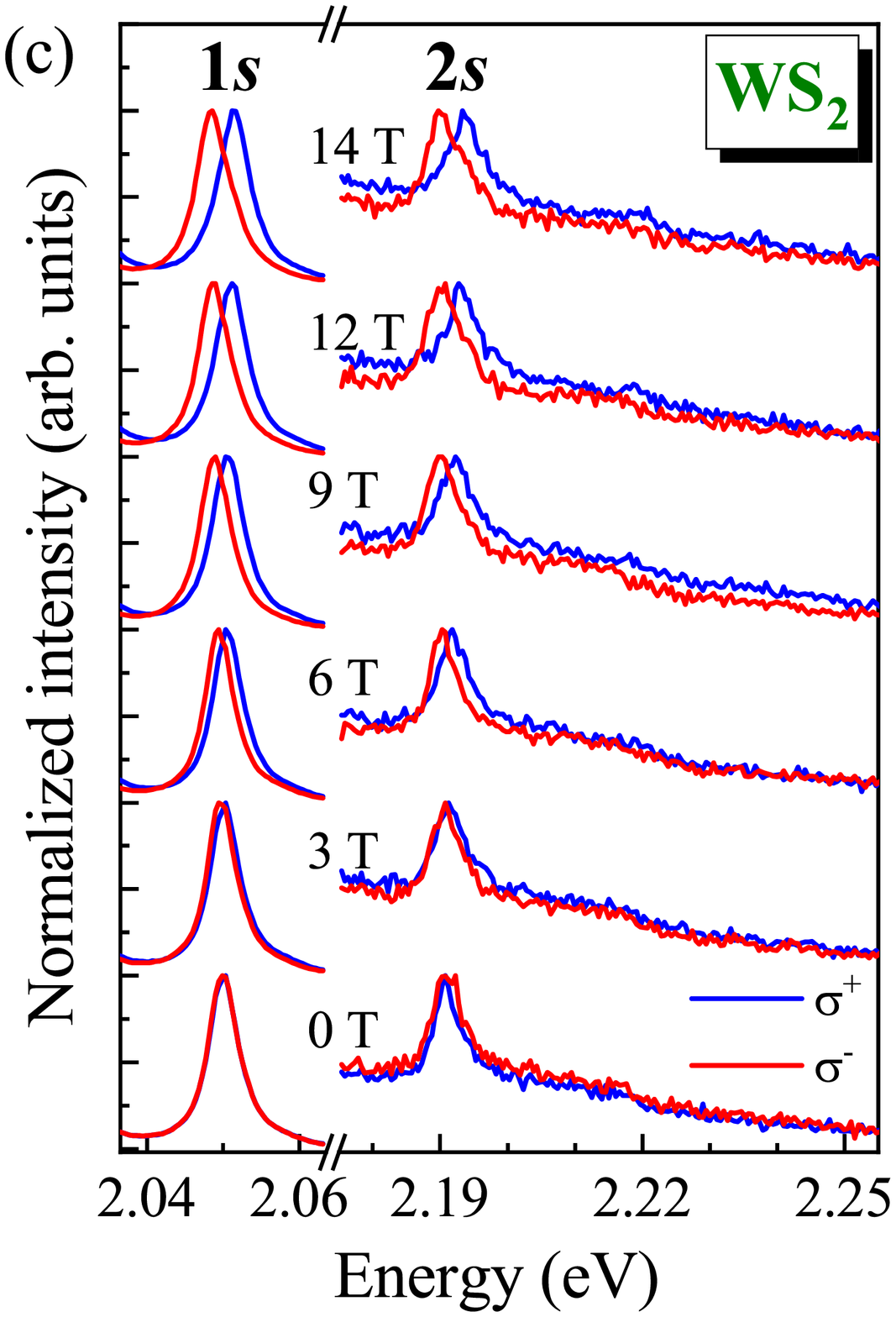}%
		\end{minipage}\hfill
		\begin{minipage}[c]{0.25\linewidth}
			\includegraphics[width=1\linewidth]{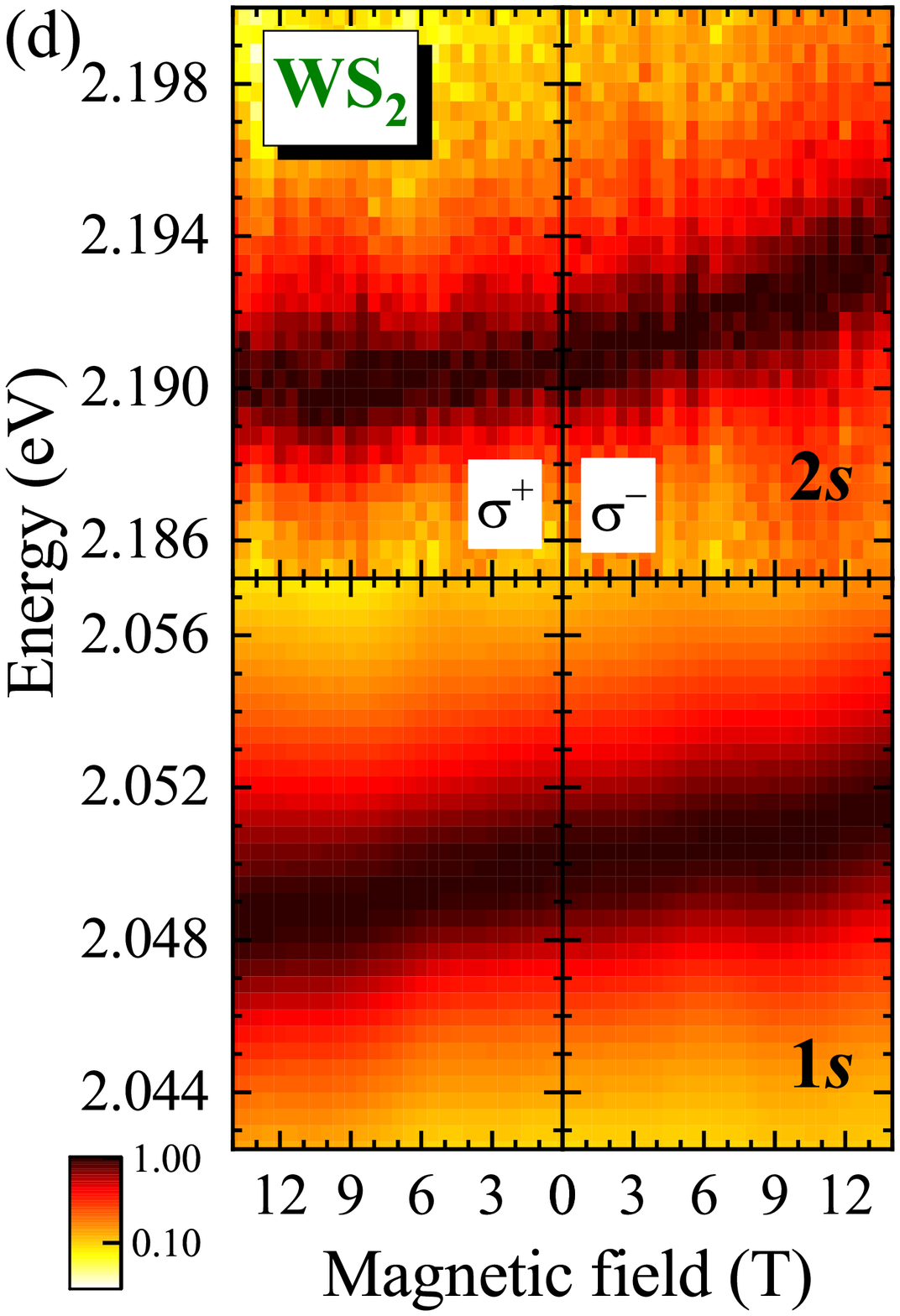}%
		\end{minipage}\hfill
		\caption{(a)/(c) Helicity-resolved ($\sigma^{\pm}$) PL spectra of MoS$_2$/WS$_2$ MLs at selected magnetic fields. The separate parts of the spectra are normalized to the intensity of the 1$s$ and 2$s$ lines. (b)/(d) False-colour map of the corresponding PL spectra from 0 to 14~T.}
		\label{fig:inne}
	\end{figure*}
\end{center}

To study the ladder of excitonic states in S-TMD monolayer, we additionally performed  the helicity-resolved magneto-photoluminescence
experiments on MoS$_2$ and WS$_2$ monolayers encapsulated in hBN, see Fig.~\ref{fig:inne}. For these materials, however, only emissions related to the
ground 1$s$  state and the first excited 2$s$ state of the A exciton is observed at whole range of applied magnetic fields. It is in contrast to
the WSe$_2$ monolayer studied in the main text, in which the emissions of higher excitonic states is better visible for bigger values of magnetic fields.

\section{D\lowercase{ependence of excitonic diamagnetic coefficients in} WS\lowercase{e}$_2$ \lowercase{monolayer}\label{sigma}}

\begin{figure}[h]
	\centering
	\includegraphics[width=0.35\linewidth]{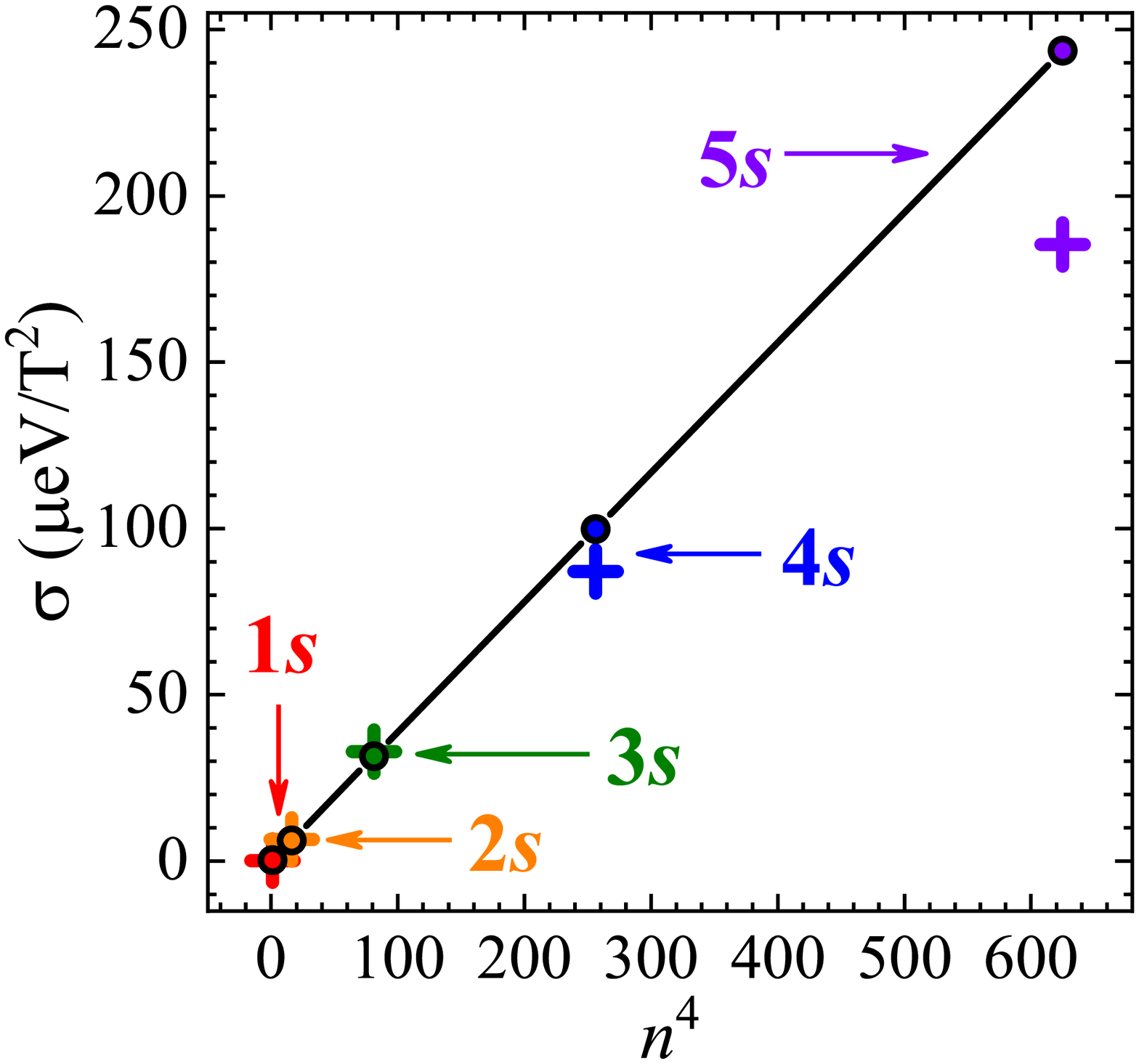}%
	\caption{Diamagnetic coefficients $\sigma$ for the excitonic states as a function $n^4$ obtained (crosses) experimentally due to the analysis performed in the main text and (circles) theoretically using Eq.~\ref{eq:diamag}. The black line connects theoretical points as a guide to the eye.}
	\label{fig:fig_diamag}
\end{figure}

To test our assumption that the excitonic spectrum in the WSe$_2$ ML encapsulated in hBN resembles
a 3D hydrogen atom ($\sim-1/n^2$), we investigate dependence of the obtained diamagnetic coefficients $\sigma$ of
excitonic states, $ns$, in this system. We found theoretically (see Eq.~\ref{eq:diamag} in the SM) that
the mean value of $r^2$ calculated with the aid of our Kratzer potential approach approximate the one for 3D hydrogen atom,
in which $r^2$ parameters of excitonic states scales with $n^4$. With the aid of Eq.~\ref{eq:diamag} and $\sigma=(e r)^2/8 \mu$, we
calculate theoretical $\sigma$ values of excitons for WSe$_2$ ML encapsulated in hBN with parameters:
$\mu$=0.2~$m_0$\cite{stierSM} and $\varepsilon=4.5$~\cite{geickSM}. The theoretical values are compared with
the experimental diamagnetic coefficients in Fig.~\ref{fig:fig_diamag}. The theoretical dependence fits very well
the experimental data up to the 4$s$ state.This additionally confirms that the excitonic spectrum of the WSe$_2$ monolayer
encapsulated in hBN corresponds to that of a 3D hydrogen atom. The apparent discrepancy between the theory and the
experiment for the 5$s$ state results in our opinion from the small range of the low-field limit, which affects the
determined $\sigma$ value for this state.

\section{L\lowercase{ow temperature reflectance contrast} \lowercase{spectra of} S-TMD \lowercase{monolayers}\label{rc}}

\begin{figure*}[h]
	\centering
	\includegraphics[width=0.75\linewidth]{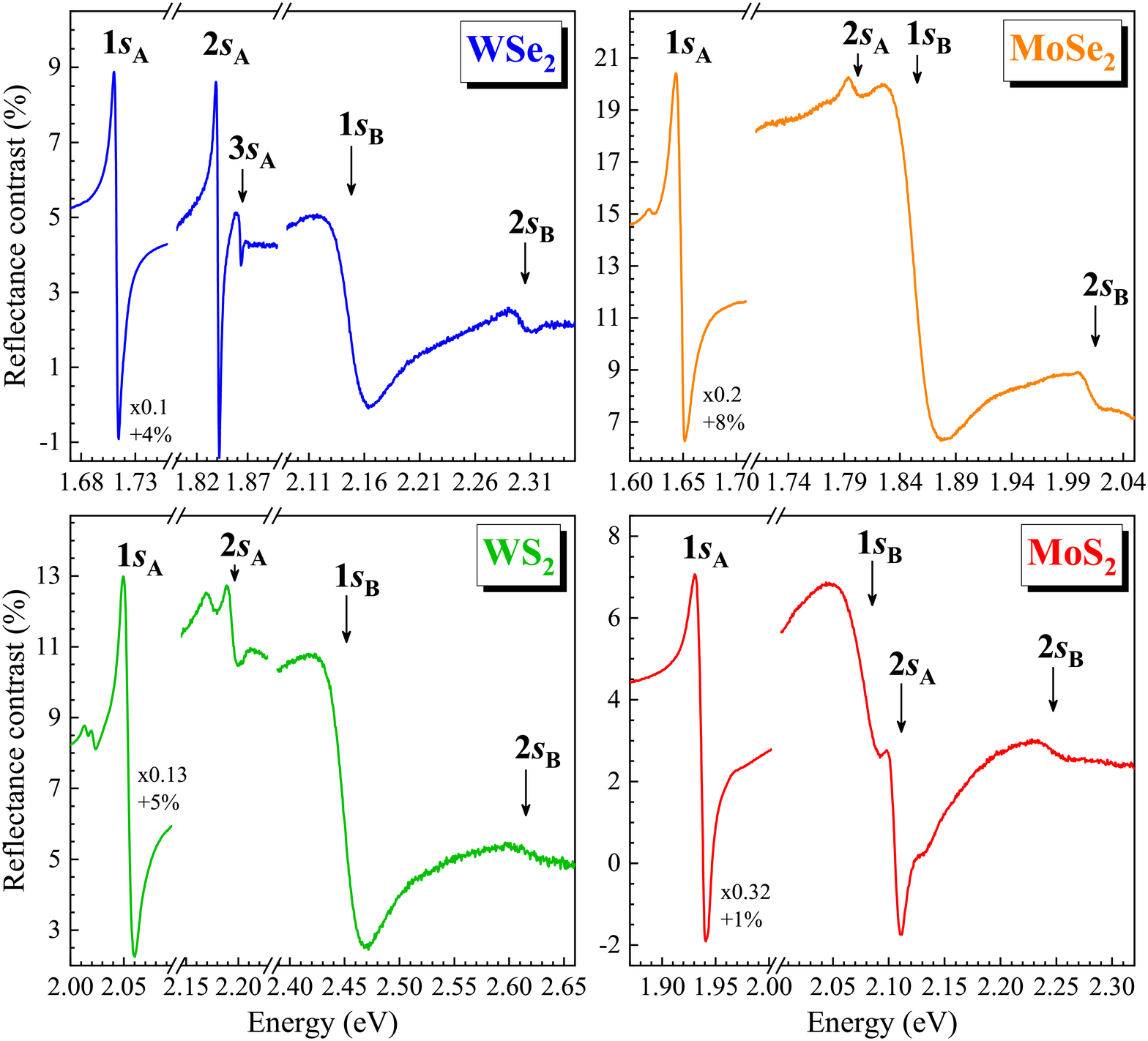}
	\caption{Low temperature RC spectra of S-TMD  monolayers measured at $T$=5~K. The spectral regions around the 1$s_{\text{A}}$ resonance are scaled and shifted for clarity.}
	\label{fig:fig_S1}
\end{figure*}

The low temperature RC spectra of WSe$_2$, MoS$_2$, WS$_2$, and MoSe$_2$ encapsulated in hBN are presented in Fig.~\ref{fig:fig_S1}.
We define the RC spectrum as $RC(E)=\frac{R(E)-R_0(E)}{R(E)+R_0(E)}\times 100\%$, with $R(E)$ and $R_0(E)$, respectively,
the reflectance of the dielectric stack composed of a monolayer encapsulated in hBN supported by a bare Si substrate and of
the two alone layers of hBN on top of Si substrate. Note that the presented spectra correspond to the PL ones shown in Fig.~4 in
the main text. First, the spectra display two pronounced resonances labelled  1$s_{\text{A,B}}$ which arise from the ground
state of the so-called A and B excitons~\cite{liSM,aroraWSe2SM,arorMoSe2SM,molasSM}. In addition to them, less pronounced
features, labelled 2$s_{\text{A,B}}$ and 3$s_{\text{A}}$, appear at about 200 meV higher in energy as compared to the 1$s_{\text{A,B}}$ ones.
The assignment of the 2$s_{\text{A}}$ and 3$s_{\text{A}}$ features to the first and the second excited state of the A exciton
is straightforward and corresponds to many other investigations on S-TMD MLs encapsulated in hBN~\cite{stierSM,robertSM,hanSM,slobodeniukSM,gerberSM}.
The origin of the 2$s_{\text{B}}$ is less clear, as it has not been reported so far. Due to the similar energy separation between 1$s_{\text{B}}$ and 2$s_{\text{B}}$ as
compared with the 1$s_{\text{A}}$ and 2$s_{\text{A}}$, we ascribed tentatively the 2$s_{\text{B}}$ features to the first excited
states of the B exciton, which, however, requires further investigations.

\newpage
\section{E\lowercase{stimation of the band-gap energy in} M\lowercase{o}S\lowercase{e}$_2$ \lowercase{monolayer}\label{bandgap}}

\begin{figure}[h]
	\centering
	\includegraphics[width=0.36\linewidth]{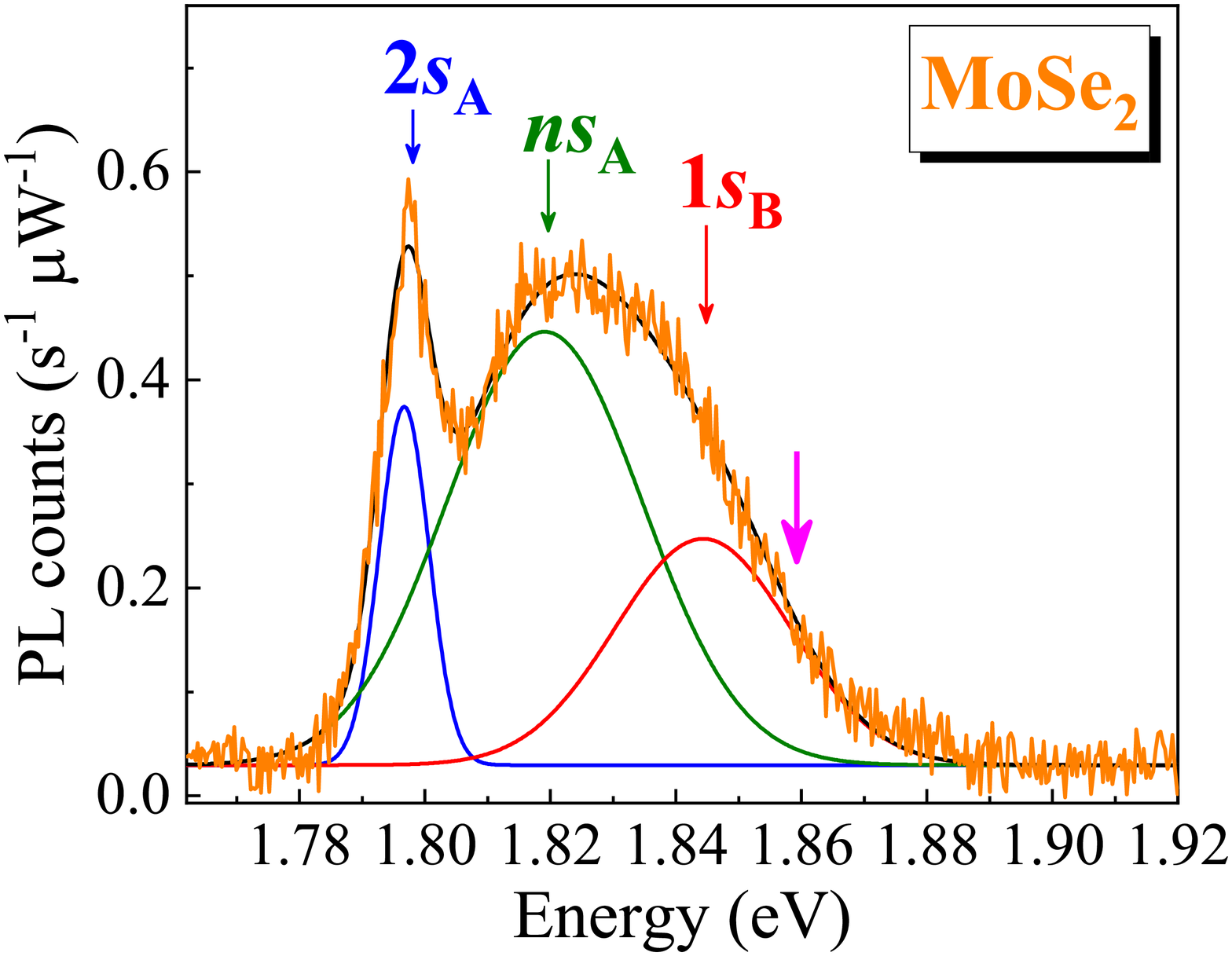}
	\caption{Low temperature photoluminescence spectrum of MoSe$_2$ monolayer at $T$=5~K limited to the high energy PL signal composed of the 2$s$ line of A exciton. The blue, green, and red curves display fits of Gaussian profiles to the corresponding 1$s_{\text{A}}$, $ns_{\text{A}}$, and 1$s_{\text{B}}$ lines. The pink vertical arrow denotes the estimated band-gap energy $E_g$.}
	\label{fig:fig_S2}
\end{figure}

As it has been discussed in the main text, the estimation of band-gap energy is essential for our analysis of excitonic
ladder in S-TMD MLs. The estimation of band-gap energy can be easily carried
out for WSe$_2$, MoS$_2$, and WS$_2$ MLs, however, this issue is more complex for the MoSe$_2$ one. This is, because PL related
to the ground state of B exciton, $1s_{\text{B}}$, appears in the spectral range of the emissions related to the $2s$ and higher $ns$
states of the A exciton, labelled as $1s_{\text{A}}$ and $ns_{\text{A}}$ in Fig.~\ref{fig:fig_S2}. To determine the band-gap
energy of MoSe$_2$ ML, we use the procedure described in the main text. The PL intensity at the band-gap energy of WSe$_2$ ML
equals 5$\%$  of the intensity of the 2$s$ exciton PL peak. In order to apply the same approach for MoSe$_2$ ML, we deconvolute
the spectrum shown in Fig.~\ref{fig:fig_S2} with three Gaussian profiles (the PL due to the $ns_{\text{A}}$ and $1s_{\text{B}}$ are
resolved spectrally). We set the linewidth of the $1s_{\text{B}}$ emission equals to 28~meV, as obtained from the upconversion PL
spectrum of MoSe$_2$ ML reported in Ref.~\citenum{hanSM}. The result of the procedure is presented in Fig.~\ref{fig:fig_S2}.
The band-gap energy in MoSe$_2$ ML determined using the procedure equals 1.861~eV and is marked in Fig.~\ref{fig:fig_S2} with pink
arrows.

\section{A\lowercase{pplication of the proposed model to the data available in the literature}\label{test}}

In the main text of this report, we have clearly demonstrated that the simple $E_{ns}=E_g-Ry^*/(n+\delta)^2$ ansatz can be well applied to reproduce the energy ladder of excitonic $s$-resonances in WSe$_2$ monolayer encapsulated in hBN. Recently, such ladders has been also extrapolated from observation of a series of excitonic resonances in very high magnetic fields,  for other, MoS$_2$, WS$_2$ and MoSe$_2$, monolayers encapsulated in hBN~\cite{gorycaSM}. Moreover, the observation of spectral series of $s$-type excitons has been also inferred from the refined analysis of the reflectance spectra in early reports on monolayer WS$_2$ deposited on Si/SiO$_2$ substrate~\cite{chernikovSM}.
\begin{figure}[h]
	\centering
	\includegraphics[width=0.92\linewidth]{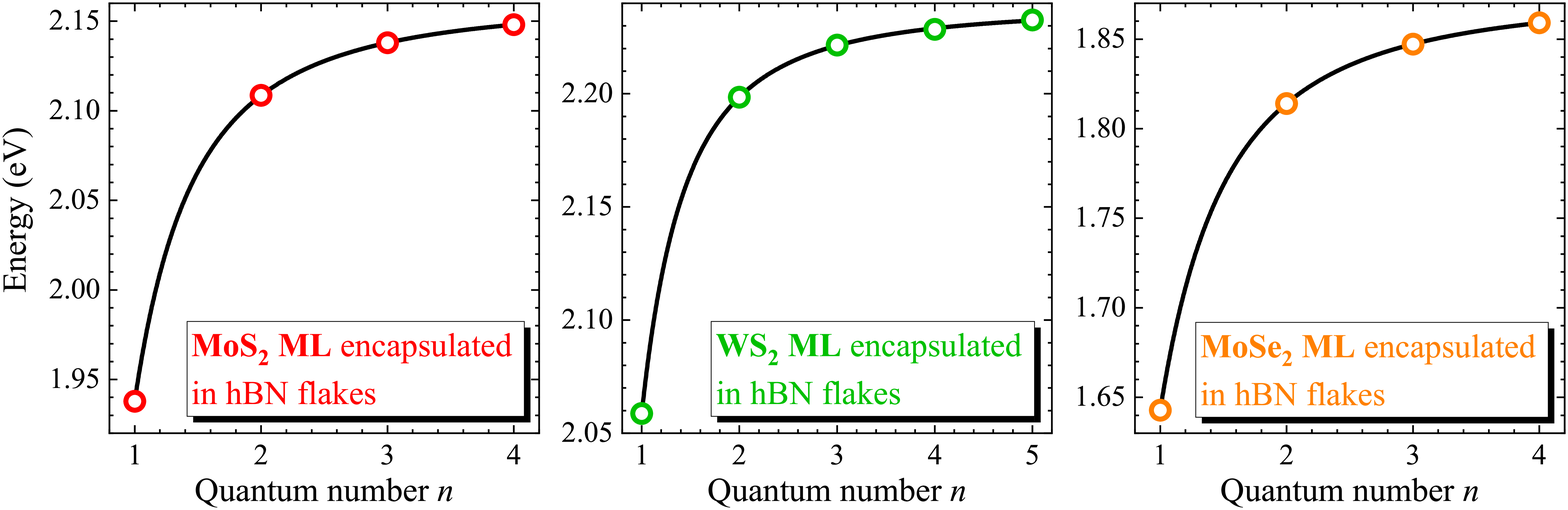}%
	\caption{Experimental transition energies for the exciton states as a function of their index, $n$, measured on S-TMD monolayers encapsulated in hBN flakes. The black curves show fits to the data with the model described by Eq.~1 in the main text. The experimental data is taken from Ref.~\citenum{gorycaSM}.}
	\label{fig:fig_S5}
\end{figure}
\begin{figure}[h]
	\centering
	\includegraphics[width=0.322\linewidth]{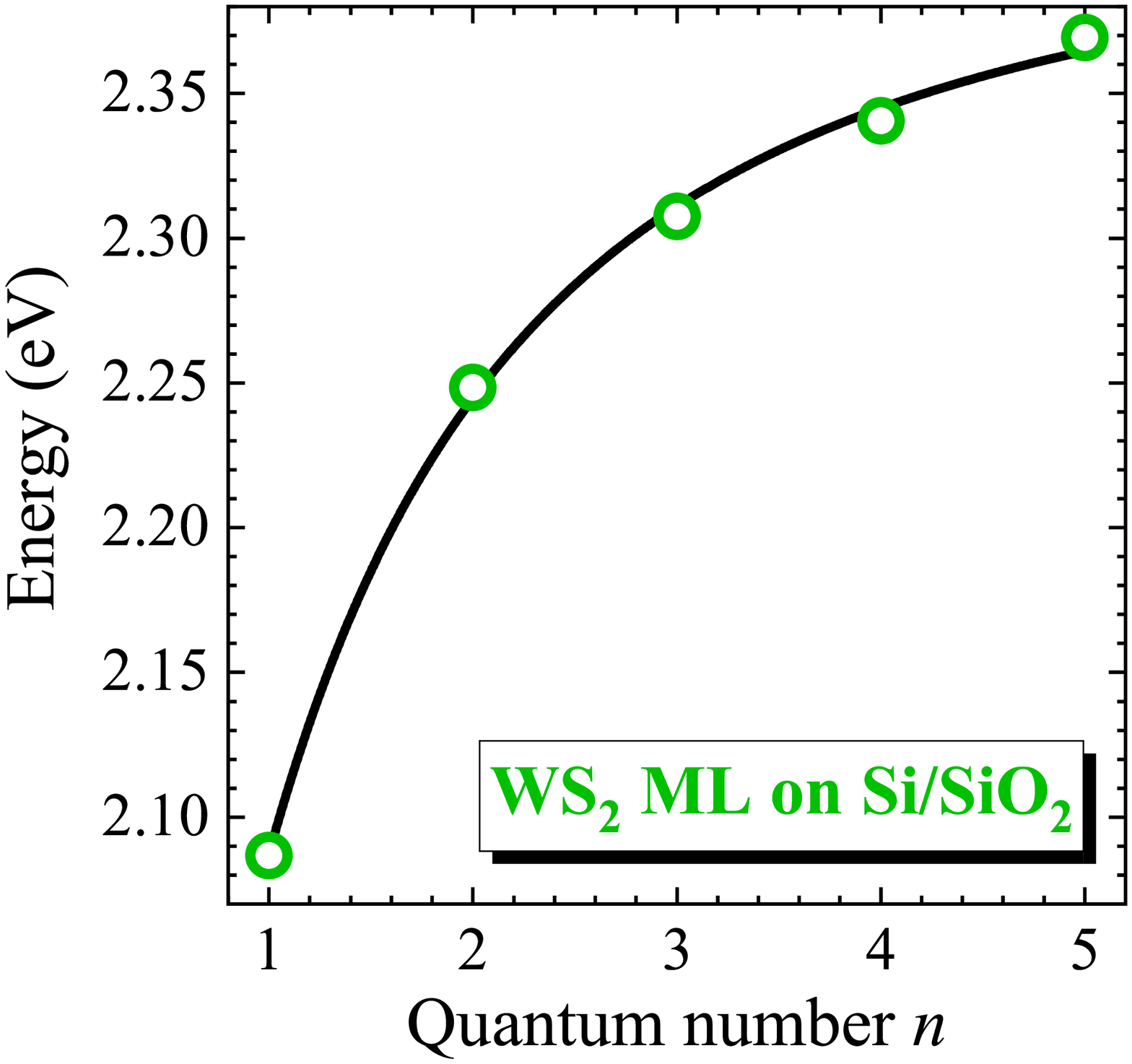}%
	\caption{Experimental transition energies for the exciton states as a function of their index, $n$, measured on WS$_2$ monolayers exfoliated onto a Si/SiO$_2$ substrate. The black curve shows a fit to the data with the model described by Eq.~1 in the main text. The experimental data is taken from Ref.~\citenum{chernikovSM}.}
	\label{fig:fig_S3}
\end{figure}

It is interesting to test the applicability of our formula against the data reported in the above references. Thus, the  energies of $s$-excitonic resonances, extracted from Refs.~\citenum{gorycaSM} and \citenum{chernikovSM}, have been fitted to the $E_{ns}=E_g-Ry^*/(n+\delta)^2$ formula, considering $E_g$, $Ry^*$, and $\delta$ as independent, adjustable parameters. As shown in Figs.~\ref{fig:fig_S5} and \ref{fig:fig_S3}, our formula fits perfectly the data for monolayers encapsulated in hBN and it is also well applied to the data for WS$_2$ on Si/SiO$_2$ system. The extracted $E_g^{\text{model}}$, $\delta^{\text{model}}$ and $E_b^{\text{model}}$ parameters
($E_b= Ry^*/(1+\delta)^2$) are listed in the Table~\ref{tab:model}. The corresponding energy values $E_g^{\text{data}}$ and $E_b^{\text{data}}$, originally reported in Refs.~\citenum{chernikovSM} and \citenum{gorycaSM}, are in perfect agreement with our findings.

According to our preceding discussion (see section \ref{sec:S3}), the applicability of our simple approach to the case of monolayers in hBN is well understood. The fact that our formula is operational for the data for WS$_2$ on Si/SiO$_2$ structure is surprising.
As a matter of fact, these data follow our formula but only in sort of the ``effective'' way. Description of these data, within our approach and notably within the Rytova-Keldysh formalism in Ref. \citenum{chernikovSM} as well, implies the use of rather unrealistic structure parameters, {\it e.g.}, the underestimated dielectric constant ($\varepsilon=1$)  and overestimated screening radius ($r_0=75\,\mbox{{\AA}}$).
\begin{center}
	\begin{table*}[t]
		\centering
		\begin{tabular}{cccccccccc}
			\hline
			\multicolumn{1}{c}{Publication} & \multicolumn{1}{|c}{S-TMD ML} & \multicolumn{1}{|c}{top medium} & \multicolumn{1}{|c}{bottom medium} &  \multicolumn{1}{|c}{$E^{\text{data}}_g$ (eV)}
			& \multicolumn{1}{|c}{$E^{\text{model}}_g$ (eV)} & \multicolumn{1}{|c}{$E^{\text{data}}_b$ (meV)} & \multicolumn{1}{|c}{$E^{\text{model}}_b$ (meV)} & \multicolumn{1}{|c}{$\delta^{\text{model}}$} \\
			\hline
			\hline
			\multicolumn{1}{c}{Ref. \citenum{gorycaSM} } & \multicolumn{1}{|c}{MoS$_2$} & \multicolumn{1}{|c}{hBN} & \multicolumn{1}{|c}{hBN} &  \multicolumn{1}{|c}{2.16} & \multicolumn{1}{|c }{2.16}  & \multicolumn{1}{|c }{221} & \multicolumn{1}{|c }{223} & \multicolumn{1}{|c}{-0.063} \\
			\multicolumn{1}{c}{Ref. \citenum{gorycaSM} } & \multicolumn{1}{|c}{WS$_2$} & \multicolumn{1}{|c}{hBN} & \multicolumn{1}{|c}{hBN} &  \multicolumn{1}{|c}{2.238} & \multicolumn{1}{|c }{2.24}  & \multicolumn{1}{|c }{180} & \multicolumn{1}{|c }{180} & \multicolumn{1}{|c}{-0.11} \\
			\multicolumn{1}{c}{Ref. \citenum{gorycaSM} } & \multicolumn{1}{|c}{MoSe$_2$} & \multicolumn{1}{|c}{hBN} & \multicolumn{1}{|c}{hBN} &  \multicolumn{1}{|c}{1.874} & \multicolumn{1}{|c }{1.87}  & \multicolumn{1}{|c }{231} & \multicolumn{1}{|c }{232} & \multicolumn{1}{|c}{0.044} \\
			\hline
\hline
			\multicolumn{1}{c}{Ref. \citenum{chernikovSM} } & \multicolumn{1}{|c}{WS$_2$} & \multicolumn{1}{|c}{air} & \multicolumn{1}{|c}{Si/SiO$_2$} &  \multicolumn{1}{|c}{2.41} & \multicolumn{1}{|c }{2.42}  & \multicolumn{1}{|c }{320} & \multicolumn{1}{|c }{328} & \multicolumn{1}{|c}{1.58} \\
			\hline
		\end{tabular}
\caption{Series of fitting parameters ($E^{\text{model}}_g$, $E^{\text{model}}_b$ and $\delta^{\text{model}}$) obtained from the analysis of data available in the literature~\cite{chernikovSM,gorycaSM} compared with the reported ones ($E^{\text{data}}_g$ and $E^{\text{data}}_b$).}
\label{tab:model}
	\end{table*}
\end{center}


\end{document}